# The Impact of Generative AI on Collaborative Open-Source Software Development: Evidence from GitHub Copilot

**Fangchen Song**
University of Texas at Austin
2110 Speedway
Austin, Texas
United States
78712
fangchen.song@mccombs.utexas.edu

**Ashish Agarwal**
University of Texas at Austin
2110 Speedway
Austin, Texas
United States
78712
ashish.agarwal@mccombs.utexas.edu

**Wen Wen**
University of Texas at Austin
2110 Speedway
Austin, Texas
United States
78712
wen.wen@mccombs.utexas.edu

## Abstract

Generative artificial intelligence (AI) facilitates content production and enhances ideation, with potentially important implications for developer productivity and participation in software development. To explore its impact on collaborative open-source software (OSS) development, we investigate the role of GitHub Copilot, a generative AI pair programmer, in OSS development where multiple distributed developers voluntarily collaborate. Using GitHub's proprietary Copilot usage data, combined with public OSS project data obtained from GitHub, we find that Copilot use increases project-level code contributions by 5.9%. This gain is accompanied by a 3.4% increase in developer coding participation and a 2.1% increase in individual code contributions. However, Copilot use is also associated with an 8% increase in coordination time and more code discussions. This reveals an important tradeoff: While AI expands who can contribute and how much they contribute, it slows coordination in collective development efforts. Despite this tension, the overall effect remains positive, resulting in a net increase in the timely merge of code contributions at the project level. Interestingly, we also find heterogeneous effects across developer roles. Peripheral developers exhibit relatively smaller increases in project-level code contributions and larger increases in coordination time than core developers. Together, our findings highlight the dual effects of AI pair programmers on code contributions and coordination in OSS development and provide implications for how generative AI may reshape the structure of OSS communities over time.

## 1. Introduction

The continuous advancements in generative artificial intelligence (AI) are transforming content production across a wide range of domains. Cutting-edge generative AI tools can not only automate mundane tasks but also enhance original content. In the context of software development, generative AI-powered pair programmers (i.e., AI pair programmers) like GitHub Copilot, Amazon Q Developer, Google Gemini, and Chat GPT can swiftly generate code based on developers' prompts, parameters, and descriptions (Dakhel et al. 2023). By reducing common coding errors and repetitive coding needs, these AI pair programmers hold great potential to reshape the software development process. A growing body of literature has investigated various impacts of generative AI on comparable, well-defined tasks on individuals, such as writing and customer service tasks (Noy and Zhang 2023, Brynjolfsson et al. 2025). However, there is limited understanding on the role of generative AI in complex tasks that involve team collaboration and require rich contextual knowledge, such as software development.

Open-source software (OSS) development has become a critical component of the global software ecosystem, generating substantial economic value. Recent estimates suggest the demand side value derived from its widespread adoption is approximately $8.8 trillion (Hoffmann et al. 2024). OSS development relies on the voluntary participation of geographically dispersed developers without formal hierarchies or standardized workflows (Levine and Prietula 2014, Lindberg et al. 2016). Developers choose to participate when the expected benefits of contributing exceed the expected costs. Expected benefits include signaling skills to the labor market, building reputation within the community, and deriving intrinsic satisfaction and a sense of achievement (Lakhani and Wolf 2003, Roberts et al. 2006, Hann et al. 2013). Expected costs primarily consist of the time and effort required to perform development tasks in a focal OSS project (Lerner and Tirole 2002).

Prior studies show that generative AI can enhance individual developer productivity, measured as the amount of code generated per developer (i.e., intensive margin), and they primarily examine organizational contexts where participation is not voluntary, and the task is well defined (Peng et al. 2023, Cui et al. 2026). However, there has been limited understanding on how generative AI affects developer participation in an OSS project, measured as the number of developers who contribute code to a focal project (i.e., extensive margin). In OSS settings where participation is entirely voluntary, ease of development may encourage developer participation. At the same time, evidence from Q&A online communities suggests that generative AI may reduce voluntary participation by diminishing opportunities for contributors to gain recognition and visibility for their expertise (Burtch et al. 2024). However, different from these communities, OSS projects involve collaborative and iterative development processes and are driven by a diverse set of intrinsic and extrinsic motivations for contribution. Given these unique features, it is not clear how generative AI affects developer participation in OSS development and the resulting project-level code contributions. To address this gap, we investigate the following research question: *RQ1. How do AI pair programmers affect project-level code contributions in OSS development?*

Beyond project-level code contributions, effective coordination is critical to the success of OSS development (Koushik and Mookerjee 1995, Reagans et al. 2016, Mawdsley et al. 2022). However, coordination is often informal and decentralized in OSS settings, shaped by fluid team composition and voluntary participation (Howison and Crowston 2014, Lindberg et al. 2016). In our study, we focus on one key aspect of coordination, namely, the coordination time needed to integrate individual code contributions into the codebase of an OSS project (Koushik and Mookerjee 1995, Reagans et al. 2016, Mawdsley et al. 2022). Shorter coordination time enables faster software development. Despite its importance, it remains unclear how AI pair programmers affect coordination time in OSS projects. Since AI pair programmers can facilitate code comprehension and evaluation, it may reduce the time required to integrate code. However, coordination time may also depend on the extent of community interactions and discussions among developers. If AI encourages greater code discussions, it may raise the overall coordination time. Thus, we investigate the following research question: *RQ2. How do AI pair programmers affect coordination time in OSS development?*

Moreover, OSS development typically involves two types of developers: core and peripheral developers (Setia et al. 2012). Core developers primarily design the overall architecture, write code, maintain control over the codebase, and guide the long-term development of the project, whereas peripheral developers contribute code on a more irregular basis and tend to focus on bug fixes and incremental feature additions (Crowston et al. 2006, Setia et al. 2012, Gousios et al. 2014). While there is overlap in activities between these two roles, core and peripheral developers play distinct roles in sustaining OSS development. Existing research on generative AI primarily examines the differential impacts based on skill levels (e.g., Brynjolfsson et al. 2025, Cui et al. 2026). However, these insights do not directly translate into understanding heterogeneous effects between core versus peripheral developers, whose differences are not necessarily driven by technical skills. Instead, these developers may differ in their expected benefits from OSS contribution and in the level of familiarity with a focal project (Espinosa et al. 2007). Project familiarity encompasses developers' contextual knowledge of the codebase, as well as their understanding of the project's historical evolution and future direction. To address this gap, we examine the following research question: *RQ3. How do AI pair programmers affect code contributions and coordination time for core versus peripheral developers differently?*

To investigate these questions, we examine how the AI pair programmer, GitHub Copilot, influences project-level code contributions and coordination time of OSS projects on GitHub, as well as how its effect varies among core versus peripheral developers. Our unit of analysis is at the project-month-level, with the sample period from January 2021 to December 2022. We use a combination of publicly available data on GitHub projects and proprietary data on Copilot use provided by GitHub organization. Our treatment group consists of projects where Copilot was both supported by local coding environments and used by developers to code. Thus, the post-treatment period includes the months during which Copilot was supported and used in a focal project, and the pre-treatment period includes all other months. The control group includes projects where Copilot was not used throughout the sample period. We estimate our model using the Generalized Synthetic Control Method (GSCM) and validate the results through different samples, matching techniques, instrumental variable estimation, alternative estimation and a range of additional robustness checks.

Our empirical results show that the use of GitHub Copilot is associated with a 5.9% increase in the number of project-level code contributions but also an 8% increase in coordination time. These findings indicate a tradeoff between contribution gains and coordination time in the OSS development following the use of Copilot. Further analysis of the underlying mechanism suggests that the observed increase in project-level code contributions is accompanied by a significant increase in both developer participation and individual productivity. At the same time, the increase in coordination time is accompanied by a higher volume of discussions surrounding code contributions, a broader set of developers participating in these discussions, and greater discussion intensity per developer. Importantly, the combined effect of these two competing forces still yields an overall positive effect on project-level timely merge of code contributions into the codebase.

Furthermore, we find that AI pair programmers lead to a greater increase in project-level code contributions made by core developers than those made by peripheral developers. This finding is consistent with our argument that core developers likely derive greater net benefits from using AI pair programmers, so they are incentivized to contribute more than peripheral developers. We also find that, following the use of AI pair programmers, there is a larger increase in coordination time for integrating code contributed by peripheral developers than core developers. This is consistent with our argument that peripheral developers' limited project familiarity may lead to more discussions on their code contributions.

Our study provides several contributions to literature. First, it contributes to the literature on generative AI in OSS development. Prior work has primarily focused on software development in organizational settings (Peng et al. 2023, Hoffmann et al. 2025, Cui et al. 2026) and shows generative AI improves individual developer productivity. However, less is known about its impact in voluntary participation in collaborative OSS development. To our knowledge, we are among the first to show generative AI encourages more developer participation in OSS development, including both coding and non-coding participation (i.e., code discussions). More importantly, we contribute to the OSS literature grounded in the cost–benefit and expectancy value framework (Lerner and Tirole 2002, Roberts et al. 2006, Stewart et al. 2006) by introducing generative AI as a novel technological factor that reshapes developers' participation decisions. Specifically, we argue that AI pair programmers simultaneously increase expected benefits by enhancing ideation capabilities and enabling exploration of alternative solutions, while reducing expected costs by lowering the time and effort required to complete development tasks. In doing so, we extend this framework by demonstrating how AI shifts contribution behavior along both the extensive margin, that is whether developers participate, and the intensive margin, that is how much they contribute.

Second, our study contributes to the literature on generative AI in team collaboration (Li et al. 2024, Dell'Acqua et al. 2025). While prior work has examined the impact of generative AI within traditional teams characterized by fixed size and formal coordination processes for performing a common task, we extend this work to open collaborative environments, where participation is voluntary and individuals perform distinct tasks that must be integrated. Importantly, our study uncovers a novel impact of generative AI on non-coding activities. Specifically, while AI tools may encourage greater developer participation in code discussions, they may also increase coordination time by requiring developers to reconcile a broader set of ideas and perspectives.

Third, our study adds to the literature that explores the heterogeneity in the impact of generative AI among individuals (Dell'Acqua et al. 2023, Demirci et al. 2025). Prior studies have focused on how individual skills play a role in shaping the effect of generative AI tools on completing discrete tasks and they found that individuals with lower skills usually obtain greater productivity gain from these tools than highly skilled individuals (e.g., Peng et al. 2023, Cui et al. 2026). In contrast, our work focuses on the unique team structure in OSS settings, namely, the existence of both core and peripheral developers, and examines how developers in these two roles may benefit differently from AI pair programmers. Unlike prior work which mostly examines the effect of AI pair programmers on reducing development costs, our study is among the first to emphasize such AI tools could increase the expected benefits of code contributions.

Our findings have important implications for OSS communities. Our results suggest that while the use of AI pair programmers can increase developer participation, it may shift the balance of contributions toward core developers, highlighting the need for governance structures that support the participation of peripheral contributors. Our findings also have implications for software development teams in firms. Developers who are more central to decision-making and system design are likely to benefit more from AI pair programmers. Additionally, as AI tools increasingly automate routine coding activities, coordination and integration may become more important constraints in software development, making organizational processes and team structures critical determinants of the returns to generative AI adoption.

## 2. Literature Review

### 2.1 Generative AI in Software Development

A growing body of literature has started to examine the impact of generative AI on software development. Most studies have focused on the implications of generative AI for individual productivity on specific tasks. For example, Peng et al. (2023) find that GitHub Copilot enables individual developers to implement an HTTP server 55.8% faster than those not using the tool. Hoffmann et al. (2025) show that GitHub Copilot causes individual developers to shift focus towards coding tasks and away from project management.

Despite these insights, research on how generative AI influences project-level outcomes for complex tasks involving multiple developers remains limited. Yeverechyahu et al. (2024) investigate the innovation capabilities of generative AI, particularly its role in extrapolative versus interpolative thinking, and compare its influence on innovation in Python versus R. Different from the existing research and built upon the OSS literature, our hypotheses are motivated by the unique characteristics of OSS, namely, the software development process in an open and collaborative environment. Because participation is often voluntary and coordination does not follow formal centralized processes in this environment, it remains unclear how generative AI influences open participation in both coding and non-coding activities, as well as team coordination.

In addition, existing research that explored heterogeneity in the roles of generative AI among individuals has mostly focused on understanding the differential effects between workers with high skills against those with low skills for comparable, well-defined tasks (e.g., Brynjolfsson et al. 2025, Cui et al. 2026). In contrast, we focus on two important roles in the OSS development: core and peripheral developers, who represent important constituents of the OSS community. Importantly the distinction between core and peripheral developers lies not in their programming skills, but in their roles, responsibilities, and the resulting level of expected benefits of code contributions and project familiarity (Crowston et al. 2006,

Setia et al. 2012). Therefore, the prediction from prior studies on how individuals with different levels of skills benefit from generative AI provides limited insights on how generative AI influences core and peripheral developers within OSS communities.

### 2.2 Generative AI and Voluntary Participation

Voluntary participation is essential to online communities. Prior research suggests that individuals contribute to online communities not only to exchange information, but also to derive social benefits (Zhang and Zhu 2011). With generative AI tools capable of producing content autonomously, there have been growing concerns that AI tools may crowd out human contributions. For example, research on platforms such as Stack Overflow finds that generative AI reduces participation by replacing straightforward information exchange (Burtch et al. 2024). Similarly, research on freelancer platforms shows a decline in job postings for simple coding tasks (Demirci et al. 2025).

However, these online communities differ from collaborative environments such as OSS development, so prior studies on the implications of AI for participation in online communities may not be held in the OSS setting. More specifically, beyond information exchange, developers participate in OSS development to solve complex problems, with the goals of improving skills, enhancing software quality, and building professional reputation (Shah 2006, Kononenko et al. 2018). On the one hand, generative AI may reduce the need for deep coding expertise, potentially diminishing developers' incentives to contribute if they no longer view OSS as a valuable avenue for skill development or reputation building. On the other hand, by helping developers overcome technical challenges more efficiently, generative AI could lower the expected costs to contribute and support the broader goals of sustaining collaborative progress. As a result, it is unclear ex ante the overall impact of generative AI on participation in OSS development.

### 2.3 Generative AI in Team Coordination

While team collaboration can improve problem-solving by leveraging diverse perspectives, it also introduces coordination challenges that can hinder team performance (Janardhanan et al. 2020, Mattarelli et al. 2022). In software development, collaboration is a core activity as developers jointly design and implement software. The extent to which developers can effectively coordinate their activities has a direct impact on software team performance (Koushik and Mookerjee 1995, Reagans et al. 2016, Mawdsley et al. 2022). In the OSS setting, team members are often globally distributed, participation is voluntary, and team composition is fluid. Without centralized task allocation or hierarchical oversight, developers must self-organize and manage interdependent tasks (Lindberg et al. 2024). This fluid and decentralized structure makes efficient coordination challenging in the OSS development environment.

Although some research has examined coordination in OSS development (Medappa and Srivastava 2019), little is known about how emerging technologies, such as generative AI, shape coordination in open collaboration. In other contexts, Li et al. (2024) demonstrate that the use of generative AI does not enhance coordination within teams, largely due to the increased volume and diversity of ideas it introduces, which can hinder convergence. Conversely, Dell'Acqua et al. (2025) find that generative AI facilitates coordination during idea generation in cross-functional teams by supporting alignment across disciplinary boundaries. In both studies, team size is fixed, and coordination is operationalized as the collective development of a single, shared solution. As such, the observed coordination primarily reflects convergence toward a common outcome, rather than the integration of heterogeneous, individual contributions. To the best of our knowledge, we are among the first to examine the impact of generative AI on team coordination in an open collaboration setting, where team size and composition are fluid and individuals are performing distinct coding tasks in a self-organizing manner. We focus specifically on coordination time, a key indicator of coordination efficiency that is critical for timely delivery of software.

## 3. Theoretical Motivation

Compared with traditional software development, OSS development exhibits several distinctive features, including voluntary participation, high developer fluidity, and limited formal authority structure (Mockus et al. 2002, Howison and Crowston 2014, Huang et al. 2026). As shown in Figure 1, OSS development usually begins with core developers designing the primary codebase and hosting it for open collaboration (Singh and Phelps 2013). Both core and peripheral developers make code contributions. With the write access and full control over the projects, core developers can submit code and either directly integrate it into the primary codebase, or have it reviewed by other core developers for integrity. Meanwhile, without the write access to codebases, peripheral developers' code needs to be reviewed by core developers, who decide whether to accept the code changes, request additional modifications, or reject them (Gousios et al. 2014, Medappa and Srivastava 2019). In the sections that follow, drawing on the cost–benefit and expectancy value model widely used in the OSS literature (Lerner and Tirole 2002, Roberts et al. 2006, Stewart et al. 2006) as well as the coordination theory (Malone and Crowston 1994, Pikkarainen et al. 2008), we examine the implications of AI pair programmers for code contributions and coordination time in light of these distinctive characteristics of OSS development.

**3.1 AI Pair Programmers and Project-level Code Contributions**

To understand how AI pair programmers influence code contributions in OSS development, we draw on the cost–benefit and expectancy value model widely used in the OSS literature (Lerner and Tirole 2002, Roberts et al. 2006, Stewart et al. 2006). According to this perspective, developers assess both expected benefits and costs they must incur for their decisions of OSS code contributions (Atkinson 1957, Hertel et al. 2003, Setia et al. 2012). Expected benefits refer to the rewards developers expect from contributing to an OSS project, including signaling skills to the labor market, building reputation within the community, and deriving intrinsic satisfaction and achievement (Lakhani and Wolf 2003, Roberts et al. 2006, Hann et al. 2013). Expected costs refer to the costs faced by developers to participate in a focal OSS project, including the time and effort required to perform OSS development tasks. These expected benefits and costs exist both at the extensive margin (i.e. developer participation) and intensive margin (i.e. individual productivity) (Lerner and Tirole 2002). Because developers have complete discretion over whether, when, and how much to contribute to the OSS development, this voluntary structure makes their decisions highly sensitive to changes in these expected benefits and costs (Wen et al. 2013).

Within this voluntary and decentralized context, AI pair programmers can shape both expected benefits and costs of code contributions. By enhancing ideation capabilities (Shaer et al. 2024), such as facilitating the generation of new features, supporting experimentation with alternative solutions, and enabling exploration of novel design approaches, AI pair programmers raise the expected benefits of code contributions, including stronger signaling of technical skills, greater reputation within the OSS community, and increased intrinsic satisfaction and achievement from producing impactful work (Lakhani and Wolf 2003, Roberts et al. 2006, Hann et al. 2013). At the same time, by reducing the time and effort required to make contributions (Peng et al. 2023), AI pair programmers lower the expected costs of contributions and make it easier for developers to allocate effort to a given project. Consistent with a cost–benefit perspective, higher expected benefits combined with lower expected costs increase the net expected benefits of code contributions. This encourages more developers to participate in an OSS project (i.e., the extensive margins) and motivates each developer to contribute more (i.e., the intensive margins), thereby leading to an increase in total project-level code contributions to the focal project.

***Hypothesis 1 (H1):*** *AI pair programmers lead to an increase in project-level code contributions in OSS development.*

### 3.2 AI Pair Programmers and Coordination Time

To understand how AI pair programmers influence coordination in OSS development, we draw on the coordination theory (Malone and Crowston 1994, Pikkarainen et al. 2008). Coordination is especially critical in interdependent task environments, where it requires continuous adjustments and information exchange (Srikanth and Puranam 2014, Oliveira and Lumineau 2017, Im and Ahuja 2023). In OSS settings, developers often work asynchronously and coordinate without formal managerial structures (Eseryel et al. 2020). The OSS development characteristics, such as voluntary participation, high developer fluidity, and limited formal authority structure, make coordination both essential for maintaining coherence and challenging to achieve (Mockus et al. 2002, Howison and Crowston 2014, Huang et al. 2026).

OSS development involves substantial interdependencies among code components and developers (Lindberg et al. 2016), making it difficult to reconcile differing perspectives during the integration process. In this study, we focus on coordination time, defined as time spent discussing code to align efforts and resolve conflicts among developers so that individual contributions can be effectively integrated into the existing codebase (Howison and Crowston 2014, Lindberg et al. 2016, Medappa and Srivastava 2019, Shaikh and Vaast 2023).

Coordination theory emphasizes group-based discussion as a key mechanism for supporting essential group functions, including information sharing and the integration of interdependent work (Malone and Crowston 1994, Pikkarainen et al. 2008). In OSS settings, online discussions serve as a primary channel to help developers clarify project goals, receive feedback on contributions, and resolve integration conflicts (Tsay et al. 2014). Accordingly, we examine developer discussions aimed at reconciling differing views among developers as a key mechanism shaping coordination time.

While encouraging code contributions, AI pair programmers may also increase discussion activities within the focal project due to the following reasons. First, by automating coding tasks, AI pair programmers free developers' time (Peng et al. 2023), enabling greater participation in discussions related to implementation and design decisions. Second, AI pair programmers assist developers in understanding algorithms and exploring alternative solutions (Barke et al. 2023), thereby increasing developers' capacity to provide feedback. Third, reductions in expected costs of code contributions may attract more developers with limited project familiarity to contribute to a focal project. Their code are likely to require more clarification, revision, or explanation, prompting additional discussions during code review.

Overall, such increased code discussions could introduce a broader set of perspectives and a greater volume of input, which makes it more difficult to align viewpoints and reach consensus. In decentralized OSS environments, where coordination relies heavily on peer discussion rather than formal authority, such increases in discussions can delay team coordination. As a result, although AI pair programmers may improve the efficiency of individual interactions and help developers respond to feedback, the expansion in discussion volume and diversity is likely to outweigh these efficiency gains. Therefore, for a given level of code contributions, the use of AI pair programmers is likely to increase coordination time before code is integrated into the project's codebase.

***Hypothesis 2 (H2):*** *AI pair programmers lead to an increase in coordination time for integrating code contributions in OSS development.*

### 3.3 Core and Peripheral Developers

The OSS development process involves two categories of developers: core and peripheral (Setia et al. 2012). Core developers, who often include the projects' administrators and key maintainers, are responsible for defining the overarching objectives and ensuring the delivery of the final software (Crowston et al. 2006). On the other hand, peripheral developers introduce additional perspectives shaped by different user needs. They mainly focus on enhancing the existing codebase such as fixing bugs and adding new features (Setia et al. 2012). Given the differences in roles and responsibilities within OSS projects between core and peripheral developers, we expect AI pair programmers to affect these two types of developers differently. In the following sub-sections, we discuss how the impact of AI pair programmers on project-level code contributions and coordination time may vary between core and peripheral developers.

#### 3.3.1. AI Pair Programmers and Code Contributions for Core vs. Peripheral Developers

As discussed above, AI pair programmers could effectively enhance developers' ideation capabilities and facilitate exploration of novel solutions, which may motivate developers to contribute more due to the increased expected benefits from code contributions. While there is some overlap in coding activities (such as adding new features and fixing bugs) between core developers and peripheral developers, the expected benefits from using AI for these tasks differ across roles. We argue that core developers are especially well positioned to realize these benefits due to the following considerations. First, core developers play a pivotal role in defining project architecture and guiding its evolution (Crowston et al. 2006). Because their work directly shapes the trajectory of the project, they have a broader scope and greater authority to implement substantive changes (Crowston et al. 2006, Setia et al. 2012). Second, core developers often possess a high level of project familiarity (i.e., contextual knowledge of the codebase, historical evolution and future direction), which allows them to evaluate AI-generated ideas and integrate them into the existing codebase more effectively. Third, core developers' contributions are highly visible and closely tied to the project's success, allowing them to obtain substantial reputational gains and skill signaling benefits (Lakhani and Wolf 2003, Roberts et al. 2006, Hann et al. 2013). Therefore, core developers may be more motivated to pursue larger, more innovative changes following the use of AI pair programmers. Overall, we expect AI pair programmers to motivate core developers to both participate more (i.e., on the extensive margin) and contribute more code individually (i.e., on the intensive margin).

In contrast, because peripheral developers often participate intermittently in project activities (Howison and Crowston 2014, Krishnamurthy et al. 2016), they tend to have lower project familiarity and to derive weaker reputational or signaling benefits from their contributions. Consequently, even when performing similar or overlapping tasks, peripheral developers face comparatively weaker incentives to leverage AI generated ideas for more substantive contributions. Thus, AI pair programmers may not increase the expected benefits of code contributions for peripheral developers to the same extent as for core developers.

On the cost side, AI pair programmers assist developers in understanding code, streamlining code generation, and simplifying implementation (Barke et al. 2023). Because core developers engage in a broader scope of code contributions, these tools are likely to reduce their total expected costs of contributing code to a greater extent, which is a function of both the volume of contributions and the marginal cost of producing each unit of code. Although AI pair programmers can substantially reduce the marginal cost of code production for peripheral developers, their relatively narrower scope of contributions limits the extent to which these reductions translate into lower total expected costs. As a result, the overall cost reduction is likely to be more pronounced for core developers than for peripheral developers.

To summarize, AI pair programmers may lead to larger increases in expected benefits of code contributions and larger reductions in total expected costs for core developers, thereby incentivizing them to contribute more than peripheral developers. As a result, we propose the following hypothesis.

***Hypothesis 3 (H3):*** *AI pair programmers lead to a larger increase in project-level code contributions made by core developers compared to peripheral developers.*

### 3.3.2. AI Pair Programmers and Coordination Time for Core vs. Peripheral Developers

As suggested above, one important reason for more code discussions following the use of AI pair programmers is that if AI pair programmers motivate developers unfamiliar with the focal project to contribute more, their code may require more clarification, revision, or explanation. We expect that the increase in coordination time could be bigger for code contributions made by peripheral developers than those made by core developers, for reasons as follows.

Peripheral developers typically engage in more bounded tasks (Setia et al. 2012) and participate intermittently in project activities (Howison and Crowston 2014, Krishnamurthy et al. 2016). As a result, a substantial share of peripheral developers is new or unfamiliar with the focal project and possess limited contextual knowledge. When AI pair programmers assist them to contribute code, their limited project familiarity means they may struggle to provide the project specific information necessary to fully leverage these tools. Although AI pair programmers can help peripheral developers generate code more quickly, the resulting code may not fully incorporate the specific architectural and strategic context of the project (Piñeiro-Martín et al. 2025). In OSS environments with flexible governance and decentralized decision making, integrating such contributions often requires additional discussion to clarify intent, resolve interdependencies, and align design choices with the broader project context (Gousios et al. 2014). As a result, at the project level, peripheral developers' contributions are on average more likely to undergo more discussions, leading to relatively longer coordination time.

In contrast, core developers are more stable participants who remain involved throughout the project lifecycle and possess deep familiarity of the project's architecture, conventions, and evolution (Dagenais and Robillard 2010). This stability and accumulated experience enable them to provide more context-rich prompts and to more effectively tailor AI-assisted outputs to the project's existing structure. Their contributions therefore tend to align more readily with existing dependencies, reducing the need for extensive clarification or discussion. Moreover, because the broader contributor base generally lacks the project familiarity needed to engage deeply in discussions surrounding these changes, core developers' contributions may attract less prolonged discussions. Consequently, the coordination time required to integrate core developers' code contributions into the codebase is likely to be relatively shorter.

Taken together, these differences in project familiarity directly affect the extent of required discussion, suggesting that AI pair programmers may have heterogeneous effects on coordination time across developer roles, with peripheral developers experiencing a relatively greater increase in coordination time than core developers.

***Hypothesis 4 (H4):*** *AI pair programmers lead to a larger increase in coordination time for integrating code contributed by peripheral developers compared to the code contributed by core developers.*

**4. Data**

We use GitHub data for our empirical analysis. GitHub is one of the largest code-hosting repositories based on the Git version control system (Dabbish et al. 2012). In this setting, a project is a fundamental unit that typically contains the source code and resource files, along with information related to the project's evolution history, high-level features, and developer details. Such projects are often used to investigate collaborative development practices (Dabbish et al. 2012). We investigate how the introduction of AI pair programmer GitHub Copilot impacts project-level code contributions and coordination time for GitHub projects. Copilot first became available in June 2021 and was followed by full public availability in June 2022[1]. Based on this timeline, we collect panel data of GitHub projects from GitHub Archive Dataset, spanning a two-year timeframe, from January 2021 to December 2022[2].

To ensure the generalizability of our analysis, we follow established procedures in the literature (Kalliamvakou et al. 2014, AlMarzouq et al. 2020) to identify active OSS projects during our panel period. We select projects with non-zero size, at least one specified programming language and license, a description, and no mirror or personal store designation. To exclude ghost or abandoned projects, we require at least one code submission every six months from 2021 to 2022 and at least one additional activity, such as a release or creation, each year. As we are interested in evaluating project-level code contributions and coordination time involving multiple developers, we focus on projects with at least three developers contributing each month. Additionally, as discussed in greater detail below, we need to use IDE[3] information to identify projects in the treatment group versus those in the control groups, so we further restrict our sample to those that disclose IDE information. These criteria result in a sample of 9,244 projects.

[1] GitHub launched the technical preview of Copilot in June 2021: https://github.blog/2021-06-29-introducing-github-copilot-ai-pair-programmer/. It then announced the formal launch and public availability of Copilot in June 2022: https://github.blog/2022-06-21-github-copilot-is-generally-available-to-all-developers/.

[2] We choose this endpoint to ensure our results are not influenced by the rise in popularity of Chat GPT, which began in early 2023.

To identify projects where Copilot was used by developers (i.e., the treatment group), we collaborated with GitHub organization, which provided proprietary aggregated Copilot usage data at the project level. This dataset indicates the monthly proportion of developers contributing code to a given project who used Copilot.[4] Since Copilot requires a compatible IDE, we also consider IDE usage. During our analysis period, only a limited number of IDEs supported Copilot: Visual Studio Code, the JetBrains suite of IDEs, Neovim, and Visual Studio.[5] To determine IDE usage, we examine the webpages of the projects in our sample to gather information on the IDE usage by contributing developers. We categorize projects with developers who used Copilot and one of the supported IDEs as our treatment group, and those using unsupported IDEs and not using Copilot as our control group. In total, our sample includes 5,687 projects in the treatment group and 3,557 projects in the control group. We designate the first month when Copilot was supported by IDEs and used by a non-zero proportion of developers as the treatment start time for each project. The months before this time are defined as the pre-treatment period and the months after this time as the post-treatment period for each project, covering the two-year span from 2021 to 2022.

To further improve comparability, we restrict the sample to project-month observations with at least one code contribution. This focus allows us to estimate Copilot's effect on collaborative development in actively maintained projects. Moreover, because the analysis of coordination time requires at least one code contribution to compute acceptance time, this criterion allows us to use the same sample to investigate H1 and H2. The final sample includes 7,637 projects, with 4,491 in the treatment group and 3,146 in the control group. Table 1 provides descriptions and summary statistics for project-month-level variables used in the main analysis. We check the robustness of our results by removing the restriction on the number of code contributions. Additionally, we test the robustness of the results using a larger sample that includes projects with as few as two developers.

[3] IDE stands for integrated development environment, which is a software application that provides local environments for coding, testing, and debugging.
[4] We do not know the identity of individual Copilot users because of privacy concerns. Although such Copilot usage is measured at the GitHub platform level, developers are likely to integrate Copilot into their workflows extensively and thus would use it for all projects where it is feasible.
[5] GitHub lists the supported IDEs for Copilot: https://github.com/features/copilot.

[insert Table 1 here]

## 5. Empirical Analyses

### 5.1 Measures

To test H1 and H2, the objective in our empirical analyses is to determine how Copilot influences project-level code contributions and coordination time. To measure project-level code contributions, we use the total number of pull requests (PRs) submitted to each project that were eventually merged[6]. PRs represent code changes submitted by developers and need further evaluation by core developers. This evaluation results in either approval, leading to merged PRs, or rejection, leading to closed but unmerged PRs. Thus, merged PRs reflect successful code changes eventually incorporated into the OSS projects and are commonly used in the literature to assess meaningful contributions by developers (Gousios et al. 2014, Kononenko et al. 2018).

To evaluate coordination time, we follow prior work in both economics (Simcoe 2012) and software engineering (Espinosa et al. 2007, Yu et al. 2015, El Mezouar et al. 2019) by measuring the duration between the submission and acceptance of each PR. Because a project may receive many PRs in a month, we then compute the average duration (i.e., average time) to merge a PR for a project in a month.

### 5.2 Model and Estimation

In our setup, projects adopt Copilot at different times after it is available. The variation in treatment timing and the larger number of treated projects compared to untreated ones complicate the application of traditional matching techniques in this context. Given these complexities, we adopt the Generalized Synthetic Control Method (GSCM) (Xu 2017), which allows for improved estimation of treatment effects by generating weighted synthetic controls that closely resemble the treated units prior to treatment. This approach combines the idea of a synthetic control (Abadie et al. 2010) with interactive fixed effects (Bai 2009). This allows us to address multiple treated units with staggered treatment times while accounting for time-varying unobservables. The GSCM shares core assumptions with the standard synthetic control method, effectively managing unobserved time-variant confounders by giving more weight to control units that mirror the pre-treatment trends of the treatment group (Xu 2017). This approach offers improved pre-treatment control and more credible parallel trends relative to traditional difference-in-differences (DID) and matching methods and has been shown to exhibit lower bias (Xu 2017, Wang et al. 2021).

[6] More specifically, we consider all PRs initiated in a month that got merged anytime between that month and December 2023 as code contributions for that month and call these as merged pull requests.

We use the following linear factor model (Bai 2009, Xu 2017) with the unit of analysis at the project-month level:

$$Y_{it} = \delta_{it} D_{it} + X_{it}'\beta + \lambda_i' f_t + \varepsilon_{it} \qquad (1)$$

where $Y_{it}$ represents the outcome variable, which is either the project-level code contributions measured by total number of merged PRs of project $i$ in month $t$, or coordination time measured by the average time taken to merge a PR of project $i$ in month $t$. We take log transformation to reduce the skewness of both variables. $D_{it}$ is the treatment indicator which equals one if project $i$ has been developed with Copilot by month $t$ and zero otherwise. The parameter of primary interest is $\delta_{it}$, which signifies the dynamic impact of Copilot on log number of merged PRs and log average time taken to merge a PR. $\delta_{it}$ is heterogeneous treatment effect and its subscripts $i$ and $t$ indicate that the estimates vary across projects and months.

In the above equation, $f_t = [f_{1t}, \ldots, f_{rt}]'$ is an $(r \times 1)$ vector of unobserved common factors associated with factor loadings $\lambda_i' = [\lambda_{i1}, \ldots, \lambda_{ir}]$. The factor component $\lambda_i' f_t$ can be expressed as $\lambda_i' f_t = \lambda_{i1} f_{1t} + \lambda_{i2} f_{2t} + \cdots + \lambda_{ir} f_{rt}$. The factor component nests a range of unobserved heterogeneities including additive unit and time fixed effects and unit-specific linear and quadratic time trends. Note that a two-way fixed effects specification is a special case of the factor component, where $r = 2$, $f_{1t} = 1$, and $\lambda_{i2} = 1$, so that $\lambda_i' f_t = \lambda_{i1} + f_{2t}$. Here, $\lambda_{i1}$ represents the project fixed effects, while $f_{2t}$ is the month fixed effects. We specify the two-way fixed effects to account for heterogeneity across projects and time, while considering other unobserved latent factors. $X_{it}'$ is the vector of observed control variables, and $\varepsilon_{it}$ is the error term with a mean of zero.

We estimate the optimal number of latent factors using a cross-validation procedure (Xu 2017). Briefly, this involves first estimating the parameters of model using the control group data only and employing a cross-validation procedure to determine the number of latent factors $r$. Next, the optimal number of factor loadings for each treated unit is estimated by minimizing the mean squared errors of the predicted treated outcomes in the pre-treatment periods. We provide a detailed explanation of the model and the estimation including the determination of the optimal latent factors in Online Appendix A. After accounting for time and project fixed effects, the optimal number of unobserved factors determined by the cross-validation technique is zero. This suggests that the fixed effects setting has effectively accounted for any unobserved time-varying characteristics (Xu 2017).

The GSCM estimator predicts the counterfactuals for treated units in the post-treatment periods using the parameter estimates obtained in the previous two steps. The causal effect of the treatment is calculated as the average treatment effect on the treated (ATT), based on the differences between the observed outcome of a treated unit $Y_{it}(1)$ and its constructed counterfactual $\hat{Y}_{it}(0)$:

$$ATT_t = \frac{1}{|\mathcal{T}|}\sum_{i\in\mathcal{T}}\left[Y_{it}(1) - \hat{Y}_{it}(0)\right] = \frac{1}{|\mathcal{T}|}\sum_{i\in\mathcal{T}} \delta_{it} \quad (2)$$

where $\mathcal{T}$ denotes the set of treated units and $|\mathcal{T}|$ represents the number of units in $\mathcal{T}$.

***Identification:*** Project fixed effects account for static unobservable differences across projects such as the characteristics of the projects or the developers that can influence code contributions to these projects and the required coordination time. Additionally, time fixed effects account for common time trends that could influence the outcomes. There may be dynamic unobservables which could influence the outcomes. GSCM synthesizes a weighted control unit that closely mirrors the data pattern of the log number of merged PRs and the log average time taken to merge a single PR during the pre-treatment period for the treated unit. The outcome variable of this synthetic control unit during the post-treatment period serves as the counterfactual prediction for the treated unit. By modeling the trend of the outcome variables, the GSCM can naturally accommodate the influence of unobservable confounders that evolve over time. Furthermore, it allows each treatment unit to have a different treatment period and can efficiently construct synthetic control units from a relatively small control sample.

To assess the validity of results, we conduct equivalence tests to examine the presence of any pre-treatment trends. The results of these tests are reported in Online Appendix B and suggest that there are no significant pre-treatment differences in the outcomes between treated and corresponding synthetic control projects. Additionally, we perform placebo tests and visualize the time-varying treatment effects across pre-treatment and post-treatment periods to ensure the robustness of our findings. Detailed results of these analyses are presented in Online Appendix C. We also validate our results using different samples, matching technique, instrumental variable analysis, alternative estimation and a comprehensive set of additional tests as described in the following sections.

## 6. Results

### 6.1 Main Results

Table 2 presents the estimated overall effects of Copilot use on project-level code contributions and coordination time. These results indicate that, following Copilot use, project-level code contributions significantly increased, but coordination time also became longer, which support H1 and H2.

More specifically, as shown in column (1), following Copilot use, the number of merged PRs of active projects increased by 5.9%. Such an increase in project-level code contributions may be driven by an increase in the expected net benefits of code contributions. By strengthening developers' ideation capabilities (Shaer et al. 2024), AI pair programmers increase the expected benefits of contribution, including skill signaling, reputation within the OSS community, and intrinsic satisfaction. At the same time, they reduce the time and effort required to contribute, lowering expected costs. Consistent with cost–benefit and expectancy value model (Lerner and Tirole 2002, Roberts et al. 2006, Stewart et al. 2006), higher expected benefits and lower expected costs increase the net expected benefits of contributing. As shown in Section 6.3.1, such an increase could exist both at the extensive margin (i.e. encouraging broader participation) and at the intensive margin (i.e. encouraging each developer to contribute more).

Column (2) reports the overall effect of Copilot on coordination time. It shows that following the use of Copilot, the average time required to merge a single PR increased by 8% on average. This observed increase in coordination time could be attributable to increased code discussions. By automating routine coding tasks, AI pair programmers free developers' time and support their engagement in implementation and design discussions. Meanwhile, AI pair programmers support developers in comprehending algorithms and exploring alternative approaches, which strengthens their ability to provide feedback. Furthermore, AI pair programmers may facilitate new and unfamiliar developers to make code contributions, which could increase the number of discussions needed to clarify those code contributions. As we explore more in Section 6.3.2, the increased discussions could exist both at the number of developers participating in discussions and at the volumes of discussions per developer.

[insert Table 2 here]

### 6.2 Robustness Checks

To further investigate the robustness of our findings, we conduct a comprehensive set of robustness checks to examine the relationship between Copilot use and the project-level code contributions and coordination time. First, to mitigate concerns about unobservable differences in development environments and selection into Copilot compatible tools, we perform a within-IDE analysis that restricts the sample to projects developed using the same Copilot supported IDE, thereby isolating variation in adoption timing rather than IDE differences. We report details and results in Online Appendix D.

Second, to address potential selection bias arising from non-random treatment assignment and staggered adoption timing, we implement a propensity score matching (PSM) combined with difference-in-differences (DID) approach that aligns treated and control projects on observable characteristics and imposes a common treatment onset, providing a more conservative estimate of the treatment effect. The detailed results are reported in Online Appendix E and qualitatively consistent with our baseline results.

Third, to address endogeneity concerns on Copilot use, we employ an instrumental variable strategy that leverages exogenous variation in developers' exposure to Copilot based on the adoption behavior of their prior collaborators outside the focal project. The detailed discussions about this instrumental variable and results are shown in Online Appendix F.

Finally, we conduct a series of additional analyses to rule out alternative explanations. Specifically, we restrict the sample to eliminate developer overlap across treatment and control projects in order to mitigate spillover effects; remove outliers to ensure that the results are not driven by extreme observations; exclude AI-related projects to address topic-specific confounds; employ alternative measures of code contributions and coordination time; control for project workload to rule out mechanical increases in coordination time; apply alternative estimation strategies, including two-way fixed effects models; and test alternative sample definitions to assess the generalizability of the findings. Detailed results are reported in Online Appendix G. Across all specifications, the results remain qualitatively consistent with our main findings, indicating that Copilot use led to an increase in the number of merged PRs as well as a longer average time to merge a PR.

### 6.3 Mechanism Analyses

#### 6.3.1 Project-level Code Contributions

In our hypothesis development for H1, we argue that AI pair programmers could increase the expected benefits of code contribution, which exist both at encouraging more developers to participate in coding and at encourage each individual developer to contribute more. To explore these potential mechanisms, we conduct the following analyses.

***Developer Coding Participation***: We measure developer's coding participation by counting the number of unique developers whose PRs have been merged into the primary codebase (Subramaniam et al. 2009, Medappa and Srivastava 2019). We conduct the analysis using the same GSCM model as discussed above with developer's coding participation as the dependent variable. The results, shown in column (1) of Table 3, indicate that Copilot increased the number of developers whose PRs have been merged by 3.4%. Meanwhile, as shown in column (1) of Table H.1 of Online Appendix H, developer coding participation is positively correlated with project-level code contributions. These results are consistent with our argument that AI pair programmers could increase developer coding participation by increasing the expected benefits and reducing the expected costs developers face when deciding whether to contribute (Atkinson 1957, Hertel et al. 2003, Setia et al. 2012, Wen et al. 2013). As a result, project-level code contributions increase.

***Individual Productivity***: We measure individual productivity using the average number of merged PRs per developer as the dependent variable (Subramaniam et al. 2009, Medappa and Srivastava 2019). We conduct the analysis using the same GSCM model as the baseline. As shown in column (2) of Table 3, Copilot use is associated with a 2.1% increase in average number of merged PRs per developer in active projects. At the same time, as shown in column (2) of Table H.1 of Online Appendix H, the volume of individual productivity is positively correlated with project-level code contributions. These results are consistent with our argument that AI pair programmers may motivate each individual developer to code more due to the increased net expected benefits of code contributions, which in turn leads to an increase in project-level code contributions.

[insert Table 3 here]

### 6.3.2 Coordination Time

In our hypothesis development for H2, we suggest that AI pair programmers may promote more discussions in the focal OSS project, which could lead to longer coordination time. To probe this potential mechanism, we analyze how AI pair programmers influence code discussions for each merged PR. Moreover, because more code discussions for each merged PR could reflect greater developer participation in code discussion, higher code discussion intensity per developer, or both, we further examine how AI pair programmers influence each of these factors.

***Code Discussion Volume***: To provide some direct evidence on this mechanism, we re-estimate our baseline model using code discussion volume as the outcome variable. We measure code discussion volume using the average number of total comments per merged PR (Oh and Jeon 2007, Gousios et al. 2014). As shown in column (1) of Table 4, Copilot use is associated with a 6.5% increase in code discussion volume per merged PR. Moreover, as shown in column (1) of Table H.2 in Online Appendix H, we observe a positive correlation between code discussion volume and coordination time. These findings support our argument that AI pair programmers may promote more code discussions, which in turn leads to longer coordination time.

***Code Discussion Participation***: Increased participation by developers in code discussion can introduce more perspectives, thereby extending the time required to reach consensus (Weber 2006, Feri et al. 2010, Yu et al. 2015, Roels and Corbett 2024). To explore whether Copilot impacts code discussion participation, we measure code discussion participation using the average number of unique developers who left comments per merged PR (Oh and Jeon 2007). Then we employ the same GSCM model using code discussion participation as the dependent variable. Column (2) of Table 4 presents the results, showing a significant 1.7% increase in code discussion participation. Meanwhile, column (2) of Table H.2 in Online Appendix H suggests a positive correlation between code discussion participation and coordination time. These results suggest that AI pair programmers could encourage broader participation in code discussions, which leads to longer coordination time.

***Code Discussion Intensity***: AI pair programmers may also affect the intensity of code discussion made by each developer on a merged PR, which could contribute to longer coordination time. To test this, we measure code discussion intensity using the average number of comments per developer per merged PR (Oh and Jeon 2007) and use it as the dependent variable based on the same GSCM model as our baseline model. As shown in column (3) of Table 4, projects using Copilot show a 5.9% increase in code discussion intensity per developer. Column (3) of Table H.2 in Online Appendix H also shows a positive correlation between code discussion intensity and coordination time. These results indicate that AI pair programmers could encourage each individual developer to participate more in code discussion, resulting in longer coordination time.

[insert Table 4 here]

***Additional Analyses on Code Discussions:*** We further examine how Copilot affects the content of code discussions among developers. Specifically, we conduct a supplementary text analysis of topic diversity using Latent Dirichlet Allocation (LDA) to extract discussion topics from 5,404,735 merged PR comments of our selected projects in the main analysis from 2021 to 2022. Details are provided in Online

Appendix I. As shown in Table I.1, Copilot increased the diversity of discussion topics per merged PR, suggesting that Copilot also expanded the range of topics being discussed, further increasing coordination time. In addition to discussion content, we examine the structure of interactions by analyzing the number of discussion cycles per merged PR, defined as the number of iterative rounds of discussion concerning the same proposed changes within a single merged PR. The results, reported in Table I.2 of Online Appendix I, indicate that Copilot increased the number of discussion cycles per merged PR, which increases overall coordination time.

***Alternative Explanation*****:** One alternative explanation for the increased coordination time is that core developers, who are responsible for reviewing and merging the code, may be allocating less effort to these activities after adopting Copilot, as they become more engaged in contributing code themselves. To investigate this possibility, we examine three measures of core developer review activity on merged PRs: (1) the total number of reviews finished by core developers, which reflects the overall volume of review work; (2) the number of unique core developers who reviewed code submissions for merged PRs, which indicates the breadth of participation; and (3) the average number of reviews per core developer, which captures review productivity of each core developer. As reported in Table 5, we find that Copilot use is associated with an increase in all three measures. Although Copilot leads to greater code contribution volume, it also appears to enhance review productivity and encourage broader participation in review among core developers. These findings suggest that, rather than reducing their review activity, developers are completing more reviews per month following Copilot use. Therefore, the increase in coordination time does not appear to be explained by core developers spending less effort reviewing the code.

[insert Table 5 here]

### 6.4 Combined Effects of Project-level Code Contributions and Coordination Time

The results in the previous sections indicate that AI pair programmers may increase both project-level code contributions and coordination time. Given the inherent tradeoff between contribution volume and the coordination effort required to manage them (Bakos and Brynjolfsson 1993), it is not yet clear how AI pair programmers shape the overall project-level timely merge of code contributions in OSS development. To provide some empirical evidence on this, we consider the number of PRs that were merged within a particular short time window as a measure of overall project-level timely merge of code contributions. For robustness, we use three time windows based on the distribution of merge times prior to Copilot's introduction: (1) the 25th percentile (1 day), (2) the median (3 days), and (3) the 75th percentile (10 days). Then, we calculate the number of PRs merged within each of the three time windows.

The results presented in Table 6 indicate that Copilot use is associated with a 3.5% increase in PRs merged within one day, a 4.1% increase in PRs merged within three days, and a 5.1% increase in PRs merged within ten days. As reported in Table H.3 of Online Appendix H, we further confirm a positive correlation between project-level code contributions and project-level timely merge of code contributions, and a negative correlation between coordination time and project-level timely merge of code contributions. These findings suggest that, despite the increased coordination time, the net effect of Copilot on project-level timely merge of code contributions is positive.

[insert Table 6 here]

***Impact on Code Quality:*** While we find a positive impact of Copilot use on timely merge of code contributions, it is equally important to understand its impact on code quality. On one hand, Copilot may reduce errors and improve the quality of code written by individual developers. On the other hand, it could potentially lower overall quality if it encourages participation from developers who are new and unfamiliar with the project. Moreover, because Copilot increases code-related discussions and extends coordination time, it remains unclear whether the longer coordination time can improve code quality. Thus, ex-ante, the effect of Copilot on quality is theoretically ambiguous.

To examine whether and how Copilot affects code quality, we include six dependent variables that are often used by prior literature to indicate quality. First, we measure the total number of issues reported in each project. Second, since issues can vary in nature and severity, we focus specifically on bug-related issues, which are more severe and widely used in the literature as a proxy for quality (Vasilescu et al. 2015). Third, we construct two normalized measures: the average number of total issues per merged PR and the average number of bug-related issues per merged PR. These normalized metrics allow us to assess quality relative to the volume of contributions. Fourth, we examine the monthly increase of stars and forks to capture downstream effects on community recognition and developer interest (Huang et al. 2026).

As reported in Table 7, we find that although Copilot use is associated with an increase in the total number of issues and bugs, the normalized measures remained statistically unchanged. These findings suggest that the increased issues and bug reports are due to a higher volume of code contributions, rather than a reduction in the quality of individual contributions.[7] Similarly, we find no significant effect on the number of monthly stars and forks. However, we acknowledge that the downstream effects of AI pair programmers on community recognition may take time to materialize. Thus, given our relatively short sample period, cautions should be taken when interpreting this set of results.

[insert Table 7 here]

### 6.5 Core and Peripheral Developers

Based on our theoretical arguments in Section 3.3, following Copilot use, both core and peripheral developers may increase their code contributions while experiencing longer coordination time. To provide supporting evidence, we construct a set of metrics capturing absolute changes in project-level code contributions and coordination time for each group. Detailed results are reported in Online Appendix K. Because GSCM generates different synthetic controls across outcomes, the magnitudes are not directly comparable across specifications. The results show that both core and peripheral developers increased their project-level code contributions. Both groups exhibit higher developer participation and individual productivity. They also experience longer coordination time accompanied by increased discussion surrounding their code contributions.

To assess the effect of AI pair programmers on the relative project-level code contributions made by core developers compared to peripheral developers (i.e., H3), we compute the proportion of merged PRs from core developers to the total number of merged PRs in each project.[8] A positive treatment estimate on this outcome variable indicates that after using Copilot, the proportion of code generated by core developers among all code merged in a project became higher, i.e. the use of AI pair programmers led to a relative increase in contribution share from core developers.

[7] We also examine whether Copilot's effects differ by project complexity and find no significant differences between simple and complex projects. Detailed discussions of the results are in Online Appendix J.

To assess the relative change in coordination time for integrating code contributed by peripheral developers compared to the code from core developers (i.e., H4), we calculate the ratio of the average time to merge peripheral developers' PRs to the overall average code-merging time in each project. A positive estimate on this outcome would indicate that PRs from peripheral developers faced relatively longer coordination time after Copilot use.

We next examine the difference in timely merge of code contributions made by peripheral versus core developers. Specifically, we compute the proportion of PRs from peripheral developers that were merged within one, three, or ten days, and use each of these as the dependent variable. Negative estimates based on these outcome variables suggest that AI pair programmers led to relatively less timely merge of code contributions for peripheral developers than for core developers.

As shown in column (1) in Table 8, the proportion of merged PRs from core developers significantly increased by 0.054 or 6.5%[9]. This seems consistent with our argument that because core developers might receive higher expected net benefits from using AI pair programmers for code contributions, there was a greater increase in code contributions made by core developers following Copilot use, as compared to peripheral developers. Additionally, as shown in column (2), peripheral developers experienced an increase of 0.085 or 5.4%[10] in their relative average merge time. Collectively, these findings suggest that Copilot led to relatively more project-level code contributions from core developers and longer coordination time for peripheral developers. The results, as well as the ones in columns (3) to (5) that use the proportion of timely merging of peripheral developers' code as the dependent variables, suggest that the project-level productivity gain from Copilot is less for peripheral developers than core developers.

---

[8] There are only two types of developers in an OSS project: core or peripheral developers. We use this measure instead of the ratio of core developers' merged PRs to peripheral developers' merged PRs because the latter measure does not allow us to incorporate cases where peripheral developers had zero merged PR in a project in a month.

[9] The mean proportion of merged PRs by core developers to the total number of merged PRs, as reported in Table 1, is 0.83. This proportion increased by 0.054, representing a 6.5% increase.

[10] The mean ratio of the average time taken to merge a PR submitted by peripheral developers to the average time taken to merge a PR submitted by all types of developers, as reported in Table 1, is 1.56. This ratio increases by 0.085, representing a 5.4% increase.

[insert Table 8 here]

Table 9 next reports the results on the mechanisms through which Copilot shapes project-level code contributions for peripheral versus core developers differently. We find that following the use of Copilot, core developers had relatively higher coding participation, as shown in column (1), and relatively higher individual productivity, as shown in column (2). These patterns are consistent with our arguments for H3 and could potentially explain the greater increase in project-level code contributions among core developers compared with peripheral developers.

To provide some suggestive evidence on how Copilot influences the expected net benefits across developers, we examine merged PRs related to new features, as these feature-based contributions are more discretionary and reflect developers' willingness to undertake tasks that can potentially deliver high benefits. Importantly, by focusing on this category of merged PRs, we compare contributions within a similar set of tasks across core and peripheral developers, thereby mitigating concerns that observed differences are driven by task heterogeneity. We compute the proportion of feature-related PRs contributed by core developers relative to all such PRs. The results, reported in column (3), show that core developers account for a larger share of feature-related PRs following the use of Copliot, consistent with our argument that, even when engaging in comparable tasks, core developers derive greater expected net benefits from using AI pair programmers.

[insert Table 9 here]

Meanwhile, as shown in Table 10, peripheral developers' code contributions received relatively greater code discussion volume, involved more developers in code discussions, and experienced greater code discussion intensity per developer. Collectively, these results could partly explain the relatively longer coordination time for peripheral developers following Copilot use and are consistent with our arguments for H4.

[insert Table 10 here]

In addition, to understand whether the increase in peripheral developers' participation in code contributions mainly comes from new or existing peripheral developers, we examine the relationship between Copilot use and the proportion of new peripheral developers relative to the total number of peripheral developers. As shown in Table K.5 in Online Appendix K, following the use of Copilot, the increase in participation is more pronounced among new peripheral developers than existing peripheral developers. This pattern seems consistent with our cost–benefit argument. Because peripheral developers typically have limited involvement in the focal project, their expected benefits from contributing are relatively stable. AI pair programmers therefore primarily affect the cost side of their participation decision. By reducing the expected participation costs more for new peripheral developers than for existing peripheral developers, AI pair programmers may lead to a larger increase in participation in code contributions among new peripheral developers.

In our hypothesis development, we argue that another potential reason that explains the difference in code contributions and coordination time between core and peripheral developers lies in their different levels of project familiarity. To provide some suggestive evidence on this argument, we conduct a series of analyses in Online Appendix L. Specifically, as shown in Table L.1 in Online Appendix L, peripheral developers exhibit lower levels of project familiarity than core developers, as measured by their working tenure within the project. Moreover, Table L.2 shows that Copilot use is associated with a relatively greater increase in the share of code contributions made by developers with high project familiarity, suggesting that project familiarity may partially account for the observed heterogeneity in code contributions. Column (2) of Table L.2 shows that Copilot use is also associated with a greater reduction in the coordination time for merging code contributed by developers with high familiarity compared to the project average. Finally, Online Appendix L also presents some qualitative evidence on the discussion content related to peripheral developers' contributions and shows that peripheral developers received a higher proportion of comments related to context clarification, which suggests that they are less familiar with the project context than core developers.

***Alternative Explanations*****:** The heterogeneous effects of Copilot on peripheral and core developers may result from other differences between peripheral and core developers, such as programming language expertise or Copilot usage. We conduct additional analyses to examine these alternative explanations and report the detailed results in Online Appendix M. Overall, as shown in Table M.1 in Online Appendix M, there are no significant differences among core and peripheral developers in terms of their programming language expertise. Furthermore, the Copilot usage seems even higher among peripheral developers as compared to core developers. These results suggest that following Copilot use, the lower project-level code contributions and longer coordination time of peripheral developers compared to core developers are not likely to be driven by lower programming skills or lower Copilot usage by peripheral developers[11].

Another important concern is that increased activity by core developers may crowd out contributions from peripheral developers, which could lead to fewer code contributions by peripheral developers as compared to core developers. To mitigate this concern, we include additional controls that capture prior coordination time within the project. As reported in Table M.2 of Online Appendix M, the results remain consistent.

In addition, to address the concern that observed increase in coordination time may simply reflect longer waiting time before code contributions are reviewed for peripheral developers, we measure coordination time as the duration from the first review to final acceptance of a code contribution, thereby excluding any waiting time before a code submission receives its initial review. This alternative measure is supported by the prior literature that shows any review prioritization is primarily captured by the initial waiting time (Yu et al. 2015). As reported in Table M.3 of Online Appendix M, both core and peripheral developers experienced increases in the duration from the first review to final acceptance, consistent with the observed increases in code discussion for both groups. Furthermore, the increase in review duration is larger for peripheral developers, which is consistent with our argument that greater amount of discussion might be needed to integrate their contributions.

[11] In a supplementary analysis, we exclude the final month of 2022, which overlaps with the initial release of Chat GPT, and re-estimate our models for core and peripheral developers. The results remain consistent with our main findings, alleviating concerns that the use of other AI tools may confound our estimates, as Chat GPT is the only generative AI tool whose availability overlaps with our panel period.

## 7. Discussion and Conclusions

The rise of AI pair programmers has garnered increasing attention due to their potential to transform software development. In this study, we explore the effects of AI pair programmers within the context of collaborative OSS development, with a focus on their impact on project-level code contributions and coordination time. Using project-level proprietary data from GitHub organization, we evaluate the impact of GitHub Copilot, an AI pair programmer, on the development outcomes of OSS projects, along with its underlying mechanisms.

Our findings indicate that Copilot use significantly increases project-level code contributions while also lengthening coordination time, highlighting a tradeoff between facilitated code generation and heightened coordination challenges in collaborative software development. Mechanism analyses reveal that the increase in project-level code contributions is accompanied by greater developer participation in coding and higher individual coding productivity. Meanwhile, the increase in project-level coordination time is accompanied by more code discussions, including greater developers participating in code discussions and higher discussion intensity per developer.

We also examine the effects of Copilot on core and peripheral developers within the OSS community, who differ in their levels of expected benefits and familiarity with the focal projects. Our analysis shows that, following Copilot use, core developers experienced a relatively larger increase in project-level code contributions and peripheral developers experienced a larger increase in coordination time. In addition, we find that peripheral developers contribute a smaller proportion of timely merge of code contributions, suggesting that the project-level productivity gain could be relatively less for peripheral developers than for core developers.

### 7.1 Implications

#### 7.1.1 Theoretical Implications

Our study offers several theoretical implications. First, our study advances existing work on the productivity effects of generative AI for individual programmers (e.g., Peng et al. 2023; Cui et al. 2026) by shifting attention to OSS projects, where participation is collaborative and voluntary in nature. Beyond this, we add to the OSS participation literature grounded in cost–benefit and expectancy-value frameworks (Lerner and Tirole 2002, Roberts et al. 2006, Stewart et al. 2006) by identifying generative AI as an important technological driver of contribution decisions. We propose that AI pair programming tools alter developers' evaluations of both benefits and costs. They expand perceived benefits by supporting creativity and facilitating the generation of alternative solutions, while simultaneously decreasing the effort and time needed to perform development tasks. Such increases in expected benefits and reduction in expected costs exist both at the extensive margin (i.e. developer participation) and intensive margin (i.e. individual productivity) (Lerner and Tirole 2002).

Second, we contribute to the emerging literature on generative AI and collaborative work (Li et al. 2024, Dell'Acqua et al. 2025) by applying coordination theory (Malone and Crowston 1994, Pikkarainen et al. 2008) to understand how AI tools shape coordination time in OSS communities. Whereas prior studies primarily examine conventional organizational teams with clearly defined roles, stable membership, and structured coordination mechanisms (e.g., Li et al. 2024; Dell'Acqua et al. 2025), we investigate OSS environments that are inherently decentralized, volunteer-driven, and characterized by fluid participation and loosely coupled collaboration. Our findings suggest that AI pair programmers not only broaden participation and facilitate idea generation but also lengthen coordination time. Specifically, AI-assisted development might increase the need for reconciliation of differing viewpoints and integration of contributions produced by a wider and more heterogeneous set of developers.

Finally, we add to the literature on heterogeneous effects of generative AI among different developers. Prior studies on generative AI generally show that lower skilled individuals benefit more from AI tools because AI reduces task completion costs (e.g., Peng et al. 2023, Cui et al. 2026). In contrast, our study highlights that the impact of AI pair programmers also depends on developers' roles and responsibilities within OSS team structures. Specifically, we find that core developers gain more from AI pair programming than peripheral developers in OSS settings. We interpret this finding through the cost–benefit perspective by showing that core developers are better positioned to realize the benefits generated by AI pair programmers, such as signaling value, ideation opportunities, and project level impact, whereas peripheral developers primarily experience cost reductions with more limited access to these benefits.

**7.1.2 Practical Implications**

Our paper has several implications for OSS communities and firms. First, AI pair programmers can increase developer participation and code generation in OSS projects, easing concerns that generative AI may reduce human contributions. However, the benefits are uneven: core developers gain more because they are more familiar with project structure and context and expect more benefits from the tools, while peripheral developers face longer coordination time that may decrease their timely contributions. While longer coordination time may be more manageable for core developers, as they can prioritize and integrate their own contributions more effectively, it imposes greater frictions for peripheral developers. Over time, these frictions could potentially discourage peripheral contributors and concentrate activity among core developers, potentially weakening the open-source ethos of broad participation and diverse perspectives. To mitigate these risks, OSS communities should consider adopting governance practices such as structured pull-request templates, AI output validation protocols, and mentoring programs to help peripheral developers better leverage AI assistance.

Our findings also have implications for software development in firms. Although corporate software development differs from OSS in that work is formally assigned and compensated, developers still vary substantially in their degree of architectural knowledge and involvement in technical decision-making. Similar to core developers in OSS projects, developers who are more central to system design and integration may be better positioned to leverage AI pair programmers for experimentation, coordination, and higher-level problem solving. In contrast, developers with narrower implementation responsibilities may primarily benefit from AI assistance in routine coding tasks. As generative AI increasingly automates lower-level programming activities, differences in contextual knowledge, architectural oversight, and decision-making responsibility may become increasingly important determinants of productivity and innovation within software teams.

Moreover, our results indicate that improvements in code generation may shift the primary constraint in software development toward coordination. Although coordination in firms is more structured than in OSS environments, increased code output can place additional demands on review, integration, and alignment processes. As a result, the net productivity effects of AI adoption are likely to depend on the capacity of these coordination mechanisms. This underscores the role of organizational processes and structures in shaping the returns to generative AI in collaborative software development.

### 7.2 Limitations and Future Work

Our study has several limitations. First, due to privacy constraints, we do not observe individual-level Copilot usage, limiting our ability to directly measure heterogeneous AI adoption behaviors, role transitions, and spillover effects in OSS development. In addition, because our data are observational, we cannot fully capture unobserved developer characteristics such as motivation, ability, or work preferences that may shape their responses to AI tools. Second, while we examine coordination outcomes such as discussion activity and coordination time, our dataset does not allow us to identify the intermediate mechanisms through which AI tools alleviate or exacerbate coordination bottlenecks. Future research could examine these mechanisms. Third, due to our panel size, we cannot assess the long-term effects of AI pair programmers as developers gain experience with these tools. Future research could examine whether the observed increases in contribution activity and coordination time persist, stabilize, or diminish over time.

Finally, our study does not examine the broader social and governance implications of AI pair programmers in OSS communities. Beyond productivity and coordination outcomes, AI-generated code may reshape collaboration norms, authorship transparency, and traditional notions of code ownership in OSS development. Because AI systems draw on large-scale public projects without explicit attribution, they may complicate credit assignment, reduce transparency regarding code provenance, and potentially erode trust in community contribution processes. In addition, increased reliance on AI-generated suggestions may contribute to skill erosion, reinforce existing coding biases embedded in training data, or introduce security vulnerabilities in sensitive components. Future research should investigate these broader organizational and community implications and develop frameworks for safe, transparent, and ethical AI use in software development.

**Acknowledgements:**

We gratefully acknowledge the guidance and constructive feedback of the editors, Dr. Bin Gu and Dr. Kevin Hong, as well as the review team and seminar participants at the Tepper School of Business and the University of Connecticut. We thank GitHub for providing proprietary data on GitHub Copilot usage, with special thanks to Peter Cihon and Kevin Xu for their support in facilitating access to the data. We also acknowledge financial support from the McCombs School of Business through the Research Excellence Grant. Earlier versions of this research were presented at the 2023 Wharton Business & Generative AI Workshop, the 34th Workshop on Information Systems and Economics (WISE) in 2023, the 2024 BizAI Conference, the 2024 Conference on Information Systems and Technology (CIST), and the 2025 INFORMS Annual Meeting.

## Tables and Figures:

Table 1: Definitions and Summary Statistics, Project-Month-Level

| Variables | Description | Obs | Mean | Std. | Min | Max |
|---|---|---|---|---|---|---|
| Merged_PR | The number of code contributions (i.e., merged PRs). | 139,329 | 23.65 | 70.59 | 1 | 3956 |
| Merge_time | The average time taken (in hours) to accept a code submission. | 139,329 | 355.63 | 956.50 | 0 | 23810 |
| MergedPR_per_dev | The average number of code contributions per developer. | 139,329 | 3.19 | 6.63 | 1 | 613 |
| Num_devs_with_mergedPR | The number of distinct developers with code contributions. | 139,329 | 6.30 | 13.77 | 1 | 575 |
| Comments_per_mergedPR | The average number of comments per code contribution. | 139,329 | 220.44 | 646.57 | 0 | 17301 |
| Comments_per_dev_per_mergedPR | The average number of comments per developer per code contribution. | 139,329 | 118.19 | 340.95 | 0 | 9757 |
| Num_devs_with_comments_per_mergedPR | The average number of developers who provided comments per code contribution. | 139,329 | 1 | 0.95 | 0 | 7 |
| Reviews | The number of total reviews of core developers. | 139,329 | 24.7 | 85.09 | 1 | 4104 |
| Reviews_per_core | The average number of reviews per core developer. | 139,329 | 3.58 | 7.67 | 1 | 810 |
| Num_core_with_reviews | The number of core developers who reviewed code submissions. | 139,329 | 3.01 | 7.92 | 1 | 186 |
| MergedPR_1D | The number of code submissions accepted within one day. | 139,329 | 14.98 | 47.95 | 0 | 2569 |
| MergedPR_3D | The number of code submissions accepted within three days. | 139,329 | 17.56 | 55.08 | 0 | 2953 |
| MergedPR_10D | The number of code submissions accepted within ten days. | 139,329 | 20.77 | 63.37 | 0 | 3431 |
| Total_issues | The number of total issues. | 139,329 | 35.36 | 114.59 | 0 | 3741 |
| Total_bugs | The number of total bug-related issues. | 139,329 | 3.84 | 18.51 | 0 | 1049 |
| Total_issues_per_PR | The average number of total issues per code contribution. | 139,329 | 1.42 | 2.36 | 0 | 201 |
| Total_bugs_per_PR | The average number of total bug-related issues per code contribution. | 139,329 | 0.14 | 0.59 | 0 | 57 |
| Stars | The monthly increase of stars. | 139,329 | 83.21 | 583.77 | 0 | 67212 |
| Forks | The monthly increase of forks. | 139,329 | 19.58 | 176.45 | 0 | 32300 |
| Core_merged_PR_share | The share of code contributions from core developers. | 118,853 | 0.83 | 0.25 | 0.002 | 1 |
| Peri_merge_time_ratio | The ratio of the average time taken (in hours) to accept a code submission from peripheral developers to the average time taken to accept a code submission from all developers. | 77,888 | 1.56 | 2.56 | 0 | 254.1 |
| Peri_mergedPR_1D_share | The share of code submissions from peripheral developers and accepted within one day. | 66,005 | 0.41 | 0.38 | 0 | 1 |
| Peri_mergedPR_3D_share | The share of code submissions from peripheral developers and accepted within three days. | 68,419 | 0.43 | 0.37 | 0 | 1 |
| Peri_mergedPR_10D_share | The share of code submissions from peripheral developers and accepted within ten days. | 71,447 | 0.46 | 0.37 | 0 | 1 |
| Core_mergedPR_per_dev_ratio | The ratio of the average code contributions per core developer to the average code contributions per developer of both types (peripheral and core). | 118,853 | 1.22 | 0.64 | 0.032 | 38.8 |
| Core_dev_with_mergedPR_share | The share of core developers with code contributions to all developers with code contributions. | 118,853 | 0.76 | 0.29 | 0.005 | 1 |
| Core_new_feature_mergedPR_ratio | The share of new feature related code contributions by core developers relative to total number of new feature related code contributions | 17,257 | 0.43 | 0.34 | 0 | 1 |
| Peri_comment_per_mergedPR_ratio | The ratio of average number of comments per peripheral developers' merged PR to average number of comments per merged PR. | 77,888 | 0.55 | 0.60 | 0 | 13.09 |
| Peri_comment_dev_per_mergedPR_ratio | The ratio of average number of developers who provided comments per peripheral developers' merged PR to average number of developers who provided comments per merged PR. | 77,888 | 0.55 | 0.52 | 0 | 2.86 |
| Peri_comment_per_dev_per_mergedPR_ratio | The ratio of average number of comments per developer per peripheral developers' merged PR to the average number of comments per developer per merged PR. | 77,888 | 0.24 | 0.46 | 0 | 8.62 |

Table 2: The Impact of Copilot on Project-Level Code Contributions and Coordination Time

| | (1)<br>Project-level Code Contributions | (2)<br>Coordination Time |
|---|---|---|
| | Log (Merged_PR) | Log (Merge_time) |
| ATT of Copilot | 0.059*** | 0.080*** |
| | (0.007) | (0.021) |
| Project FE | Yes | Yes |
| Month FE | Yes | Yes |
| # of projects | 7,637 | 7,637 |
| Observations | 139,329 | 139,329 |

Notes: All estimations are based on the GSCM method. Robust standard errors clustered at project level in parentheses. Please refer to Table 1 for detailed definitions of the dependent variables used in this table. *** p<0.01, ** p<0.05, * p<0.1.

Table 3: The Impact of Copilot on Project-Level Code Contributions (Mechanism)

| | (1)<br>Developer Coding Participation | (2)<br>Individual Code Contributions |
|---|---|---|
| | Log (Num_devs_with_mergedPR) | Log (MergedPR_per_dev) |
| ATT of Copilot | 0.034*** | 0.021*** |
| | (0.004) | (0.005) |
| Project FE | Yes | Yes |
| Month FE | Yes | Yes |
| # of projects | 7,637 | 7,637 |
| Observations | 139,329 | 139,329 |

Notes: All estimations are based on the GSCM method. Robust standard errors clustered at project level in parentheses. Please refer to Table 1 for detailed definitions of the dependent variables used in this table. *** p<0.01, ** p<0.05, * p<0.1.

Table 4: The Impact of Copilot on Project-Level Coordination Time (Mechanism)

| | (1)<br>Discussion Volume | (2)<br>Developer Discussion Participation | (3)<br>Individual Discussion Intensity |
|---|---|---|---|
| | Log (Comments<br>_per_mergedPR) | Log (Num_devs_with_comments<br>_per_mergedPR) | Log (Comments_per_dev<br>_per_mergedPR) |
| ATT of Copilot | 0.065*** | 0.017*** | 0.059*** |
| | (0.017) | (0.005) | (0.014) |
| Project FE | Yes | Yes | Yes |
| Month FE | Yes | Yes | Yes |
| # of projects | 7,637 | 7,637 | 7,637 |
| Observations | 139,329 | 139,329 | 139,329 |

Notes: All estimations are based on the GSCM method. Robust standard errors clustered at project level in parentheses. Please refer to Table 1 for detailed definitions of the dependent variables used in this table. *** p<0.01, ** p<0.05, * p<0.1.

Table 5: The Impact of Copilot on Core Developers' Review Activities

| | (1)<br>Total Reviews | (2)<br>Individual Review Productivity | (3)<br>Developer Review Participation |
|---|---|---|---|
| | Log (Reviews) | Log (Reviews_per_core) | Log (Num_core_with_reviews) |
| ATT of Copilot | 0.059*** | 0.016** | 0.036*** |
| | (0.007) | (0.007) | (0.003) |
| Project FE | Yes | Yes | Yes |
| Month FE | Yes | Yes | Yes |
| # of projects | 7,637 | 7,637 | 7,637 |
| Observations | 139,329 | 139,329 | 139,329 |

Notes: All estimations are based on the GSCM method. Robust standard errors clustered at project level in parentheses. Please refer to Table 1 for detailed definitions of the dependent variables used in this table. *** p<0.01, ** p<0.05, * p<0.1.

Table 6: The Impact of Copilot on Project-Level Timely Merge of Code Contributions

| | (1) Timely Merged PRs (1 Day) | (2) Timely Merged PRs (3 Days) | (3) Timely Merged PRs (10 Days) |
|---|---|---|---|
| | Log (MergedPR_1D) | Log (MergedPR_3D) | Log (MergedPR_10D) |
| ATT of Copilot | 0.035*** | 0.041*** | 0.051*** |
| | (0.009) | (0.009) | (0.008) |
| Project FE | Yes | Yes | Yes |
| Month FE | Yes | Yes | Yes |
| # of projects | 7,637 | 7,637 | 7,637 |
| Observations | 139,329 | 139,329 | 139,329 |

Notes: All estimations are based on the GSCM method. Robust standard errors clustered at project level in parentheses. Please refer to Table 1 for detailed definitions of the dependent variables used in this table. *** p<0.01, ** p<0.05, * p<0.1.

Table 7: The Impact of Copilot on Project-Level Code Quality

| | (1) Log (Total_issues) | (2) Log (Total_bugs) | (3) Log (Total_issues_per_PR) | (4) Log (Total_bugs_per_PR) | (5) Log (Stars) | (6) Log (Forks) |
|---|---|---|---|---|---|---|
| ATT of Copilot | 0.070*** | 0.052* | 0.006 | 0.002 | -0.008 | 0.008 |
| | (0.007) | (0.030) | (0.004) | (0.002) | (0.010) | (0.006) |
| Project FE | Yes | Yes | Yes | Yes | Yes | Yes |
| Month FE | Yes | Yes | Yes | Yes | Yes | Yes |
| # of projects | 7,637 | 7,637 | 7,637 | 7,637 | 7,637 | 7,637 |
| Observations | 139,329 | 139,329 | 139,329 | 139,329 | 139,329 | 139,329 |

Notes: All estimations are based on the GSCM method. Robust standard errors clustered at project level in parentheses. Please refer to Table 1 for detailed definitions of the dependent variables used in this table. *** p<0.01, ** p<0.05, * p<0.1.

Table 8: Heterogeneous Effect of Copilot on Core versus Peripheral Developers

| | (1) Project-level Code Contributions | (2) Coordination Time | (3) Timely Merged PRs (1 Day) | (4) Timely Merged PRs (3 Days) | (5) Timely Merged PRs (10 Days) |
|---|---|---|---|---|---|
| | Core_merged_ PR_share | Peri_merge_ time_ratio | Peri_mergedPR _1D_share | Peri_mergedPR _3D_share | Peri_mergedPR _10D_share |
| ATT of Copilot | 0.054*** | 0.085** | -0.021*** | -0.024*** | -0.020*** |
| | (0.017) | (0.036) | (0.006) | (0.006) | (0.005) |
| Project FE | Yes | Yes | Yes | Yes | Yes |
| Month FE | Yes | Yes | Yes | Yes | Yes |
| # of projects | 6,423 | 4,154 | 3,350 | 3,528 | 3,729 |
| Observations | 118,853 | 77,888 | 66,005 | 68,419 | 71,447 |

Notes: This table is based on a smaller sample than the main sample because this analysis requires at least one merged PR from core or peripheral developers (in order to compute merge time). The sample becomes smaller in columns (3) to (5) due to the exclusion of observations with zero PR merged in that particular time window, which would lead to a value of zero for the denominator. All estimations are based on the GSCM method. Robust standard errors clustered at project level in parentheses. Please refer to Table 1 for detailed definitions of the dependent variables used in this table. *** p<0.01, ** p<0.05, * p<0.1.

Table 9: Heterogeneous Effect of Copilot on Core versus Peripheral Developers

(Code Contribution Mechanism)

| | (1) Developer Coding Participation | (2) Individual Productivity | (3) Expected Benefits (Core to All) |
|---|---|---|---|
| | Core_dev_with_mergedPR_share | Core_mergedPR_per_dev_ratio | Core_new_feature_mergedPR_ratio |
| ATT of Copilot | 0.023*** | 0.031*** | 0.023** |
| | (0.007) | (0.011) | (0.011) |
| Project FE | Yes | Yes | Yes |
| Month FE | Yes | Yes | Yes |
| # of projects | 6,423 | 6,423 | 837 |
| Observations | 118,853 | 118,853 | 17,257 |

Notes: All estimations are based on the GSCM method. Robust standard errors clustered at project level in parentheses. Please refer to Table 1 for detailed definitions of the dependent variables used in this table. *** p<0.01, ** p<0.05, * p<0.1.

Table 10: Heterogeneous Effect of Copilot on Core versus Peripheral Developers (Coordination Time Mechanism)

| | (1)<br>Discussion Volume | (2)<br>Discussion Participation | (3)<br>Individual Discussion Intensity |
|---|---|---|---|
| | Peri_comment_<br>per_mergedPR_ratio | Peri_comment_dev_<br>per_mergedPR_ratio | Peri_comment_<br>per_dev_per_mergedPR_ratio |
| ATT of Copilot | 0.029*** | 0.026*** | 0.013** |
| | (0.008) | (0.006) | (0.006) |
| Project FE | Yes | Yes | Yes |
| Month FE | Yes | Yes | Yes |
| # of projects | 4,154 | 4,154 | 4,154 |
| Observations | 77,888 | 77,888 | 77,888 |

Notes: All estimations are based on the GSCM method. Robust standard errors clustered at project level in parentheses. Please refer to Table 1 for detailed definitions of the dependent variables used in this table. *** p<0.01, ** p<0.05, * p<0.1.

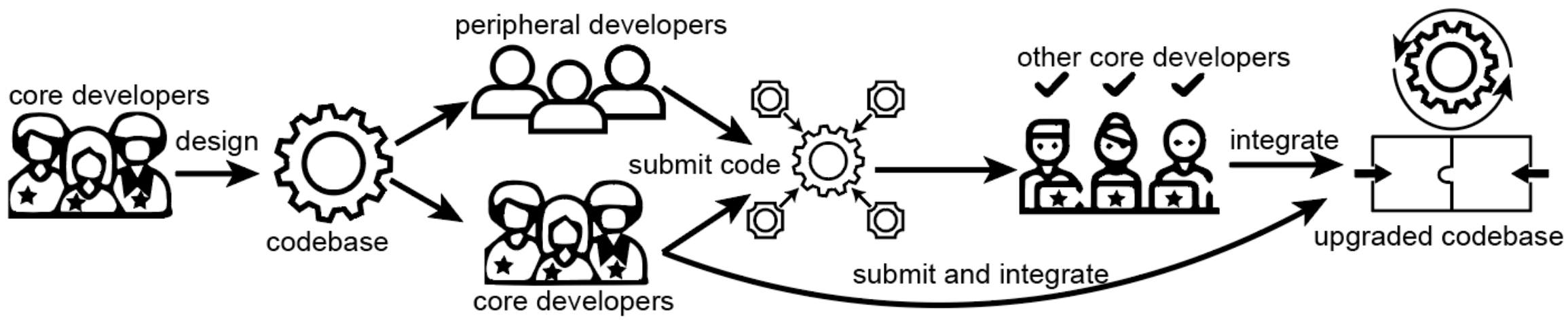

Figure 1: Open-Source Software Development Process

# ONLINE APPENDIX

## Appendix A: Estimation Details of the Generalized Synthetic Control Method (GSCM)

Our main causal inference uses the GSCM (Xu 2017). Below, we provide a brief overview of its estimation process for our study.

**Model**: The GSCM combines the Interactive Fixed Effects (IFE) model (Bai 2009), which considers latent factors with effects that can vary across time and units, with the Synthetic Control (SC) method (Abadie et al. 2015), which creates synthetic control units to act as counterfactuals for the treated units. This combination enables GSCM to (1) relax the assumption of pre-treatment parallel trends, and (2) consider potential unobserved factors that vary over time at the unit level. Furthermore, GSCM offers several advantages over other estimation methods. Unlike the SC method, which is limited to a single treated unit, GSCM is designed for multiple treated units, eliminating the need to construct synthetic control units for each treatment unit one by one. Additionally, unlike the typical difference-in-differences model combined with a specific matching technique, which is most effective for a single treatment turn-on time and a large control group, GSCM allows each treatment unit to have a different treatment period and can efficiently construct synthetic control units from a relatively small control sample. These advantages make GSCM particularly well-suited to our data, as projects adopted Copilot at different points in time and we have relatively fewer projects in the control sample than in the treatment sample.

First, the GSCM adopts a linear IFE framework to model the latent factors. In our study, the outcome $Y_{it}$, the log number of merged pull requests (PR) and the log average time taken to merge a single PR of project $i$ in month $t$, can be expressed as:

$$Y_{it} = \delta_{it} D_{it} + X_{it}'\beta + \lambda_i' f_t + \varepsilon_{it} \qquad (1)$$

where $D_{it}$ is the treatment indicator which equals one if project $i$ has been developed with Copilot by month $t$ and zero otherwise; the parameter of primary interest is $\delta_{it}$, which signifies the dynamic impact of Copilot on the log number of merged PRs and the log average time taken to merge a single PR. $\delta_{it}$ is the heterogeneous treatment effect and its subscripts $i$ and $t$ indicate that the estimates vary across units and time. Let $r$ be the number of latent factors, then $f_t$ represents the $(r \times 1)$ vector of unobserved common factors, and $\lambda_i'$ is the $(1 \times r)$ vector of unknown factor loadings. The factor component $\lambda_i' f_t$ can be expressed as $\lambda_i' f_t = \lambda_{i1} f_{1t} + \lambda_{i2} f_{2t} + \cdots + \lambda_{ir} f_{rt}$. A two-way fixed effects specification is a special case of the factor component, where $r = 2$, $f_{1t} = 1$, and $\lambda_{i2} = 1$, so that $\lambda_i' f_t = \lambda_{i1} + f_{2t}$. Here, $\lambda_{i1}$ represents the project fixed effects, while $f_{2t}$ is the month fixed effects. We specify the two-way fixed effects to control for heterogeneity across projects and time, while considering other unobserved latent factors. In general, $\lambda_i' f_t$ is estimated by an iterative factor analysis of the residuals from the model. If optimal number of unobserved latent factors determined by the cross-validation technique is zero, it means that the fixed effects setting has effectively accounted for any unobserved time-varying characteristics (Xu 2017). Lastly, $\varepsilon_{it}$ is the error term with a mean of zero.

**Estimation of Latent Factors**: An essential task when utilizing the IFE framework is to identify the number of latent factors, denoted as $r$. Xu (2017) suggests a predictive analytics method for this purpose. For clarity, we divide the total number of projects into two groups: $Tr$ for treated projects and $Co$ for control projects. The latent-factor selection algorithm first applies Equation (1) to data from control units only, covering both pre-treatment and post-treatment time periods, using the following equation:

$$Y_{it} = X_{it}' \beta + \lambda_i' f_t + \varepsilon_{it}, \forall i \in Co \qquad (2)$$

Since the control projects never received treatment during the data-collection period, $\delta_{it} D_{it}$ is excluded from Equation (2). The value of $r$ can vary within a range of candidate values specified by the researchers. For each specified $r$, the algorithm runs Equation (2) and derives the estimates $\hat{\beta}$ and $\hat{f}_t$. Subsequently, a cross-validation procedure is carried out for all treated projects. Specifically, let $s$ be the index of a pre-treatment period, ranging from one (the first pre-treatment period) to $t^0$ (the last pre-treatment period). Equation (3) is then applied to each pre-treatment period, starting with $s = 1$, for all projects in the treatment group by leaving one period out and using the remaining periods to estimate the factor loadings.

$$\hat{\lambda}_{i,-s} = (\hat{f}_{-s}^{0\prime} \hat{f}_{-s}^{0})^{-1} \hat{f}_{-s}^{0\prime} (Y_{i,-s}^{0} - X_{i,-s}^{0\prime} \hat{\beta}), \; i \in Tr, \; s = 1, \ldots, t^0 \qquad (3)$$

In Equation (3), $\hat{f}_{-s}^{0}$ and $\hat{\beta}$ are the estimates obtained from Equation (2). The superscript 0 indicates periods before the introduction of the treatment, and the subscript $-s$ denotes all periods except for $s$. Based on these estimates, the algorithm predicts the outcome for treated unit $i$ in period $s$ using $\hat{Y}_{is}(0) = X_{is}'\hat{\beta} + \lambda_{i,-s}'\hat{f}_s$ and records the out-of-sample error $e_{is} = Y_{is}(0) - \hat{Y}_{is}(0)$ for all $i \in Tr$. In the final step, the algorithm calculates the Mean Squared Error (MSE) of the prediction, summed over all pre-treatment periods $s$, for each candidate value of $r$. The value of $r$ that minimizes this prediction error is selected, and the corresponding set of latent factors will be used in the causal inference process.

**Estimation of Average Treatment Effects**: The GSCM algorithm uses the estimated function to predict the counterfactuals for treated units, denoted as $\hat{Y}_{it}(0)$, representing the outcome of treated projects had they not received treatment in the post-treatment period. The causal effect of the treatment is quantified as the Average Treatment effect on the Treated unit (ATT), calculated mathematically as:

$$ATT_t = \frac{1}{|\mathcal{T}|}\sum_{i\in\mathcal{T}}\left[Y_{it}(1) - \hat{Y}_{it}(0)\right] \text{ for } t > t^0 \qquad (4)$$

where $\mathcal{T}$ denotes the set of treated units and $|\mathcal{T}|$ represents the number of units in $\mathcal{T}$. $Y_{it}(1)$ is the observed outcome for treated project $i$ at time $t$. The essence of GSCM lies in the fact that if the predicted outcomes for the treated projects during the pre-treatment periods are accurate, the algorithm can produce a valid counterfactual for each treated unit during the post-treatment periods. Finally, the GSCM uses bootstrapping to estimate confidence intervals and standard errors (Xu 2017).

Unlike typical econometric modeling, the latent-factor selection algorithm in GSCM prioritizes models with the lowest out-of-sample prediction error to accurately predict counterfactuals, rather than those with the best fit based on information criterion. As a result, models with more latent factors may not be selected, even if they account for more confounding factors. Moreover, the GSCM without any latent factor still yields a more valid estimate than a difference-in-differences (DID) model, except under two conditions: (1) the treatment is randomly assigned, satisfying the parallel trend assumption; and (2) no treatment heterogeneity exists. If either of these conditions is not met, the DID estimate is likely to be invalid.

## Appendix B: Equivalence Test of Pre-trend in GSCM

We apply the equivalence test to examine the presence of a pre-treatment trend in GSCM (Pan and Qiu 2022, Egami and Yamauchi 2023, Wang et al. 2023). This is a standard way to test the performance of GSCM (Liu et al. 2024), as the equivalence test better incorporates substantive considerations of what constitutes good balance on covariates and placebo outcomes compared to traditional tests (Hartman and Hidalgo 2018). Specifically, we use the two-one-sided $t$ (TOST) test. The test includes an equivalence range within which differences are deemed inconsequential (Hartman and Hidalgo 2018, Lakens et al. 2018). The test is considered passed if the average prediction error for any pre-treatment period is within the equivalence range (Liu et al. 2024). Hartman and Hidalgo (2018) recommend calibrating equivalence range based on the standard deviation of the outcome when standardized benchmarks are unavailable. Building on this idea, Liu et al. (2024) propose a benchmark of ±0.36 times the standard deviation of the residualized untreated outcome, and this practice has been adopted in subsequent studies examining topics such as mobile app adoption and investor behavior as well as the impact of ride sharing platforms on public transit systems (Pan and Qiu 2022, Liu et al. 2025). Importantly, the implementation we use follows this logic but allows the threshold to vary with the data and estimation uncertainty. As a result, the implied equivalence range in our setting correspond to approximately 0.14 to 0.26 times the standard deviation of the residualized untreated outcome across different outcomes, which are more conservative than the commonly used 0.36 benchmark.

The result of the equivalence test for the main outcome variables, including project-level code contributions, coordination time and overall project-level timely merge of code contributions which combines the effect of these two competing forces, are shown in Figures B.1, through B.5. We observe that the average prediction error with 90% confidence intervals (the gray dotted line) is within the equivalence range (the red dotted line). Thus, we can reject the null of inequivalence (equivalence test p-value is 0) and conclude that there exists no pre-treatment trend (Hartman and Hidalgo 2018, Pan and Qiu 2022, Egami and Yamauchi 2023, Wang et al. 2023, Liu et al. 2024). This indicates that a sufficient set of confounders have been controlled to address the endogeneity concerns and that GSCM provides a good control group.

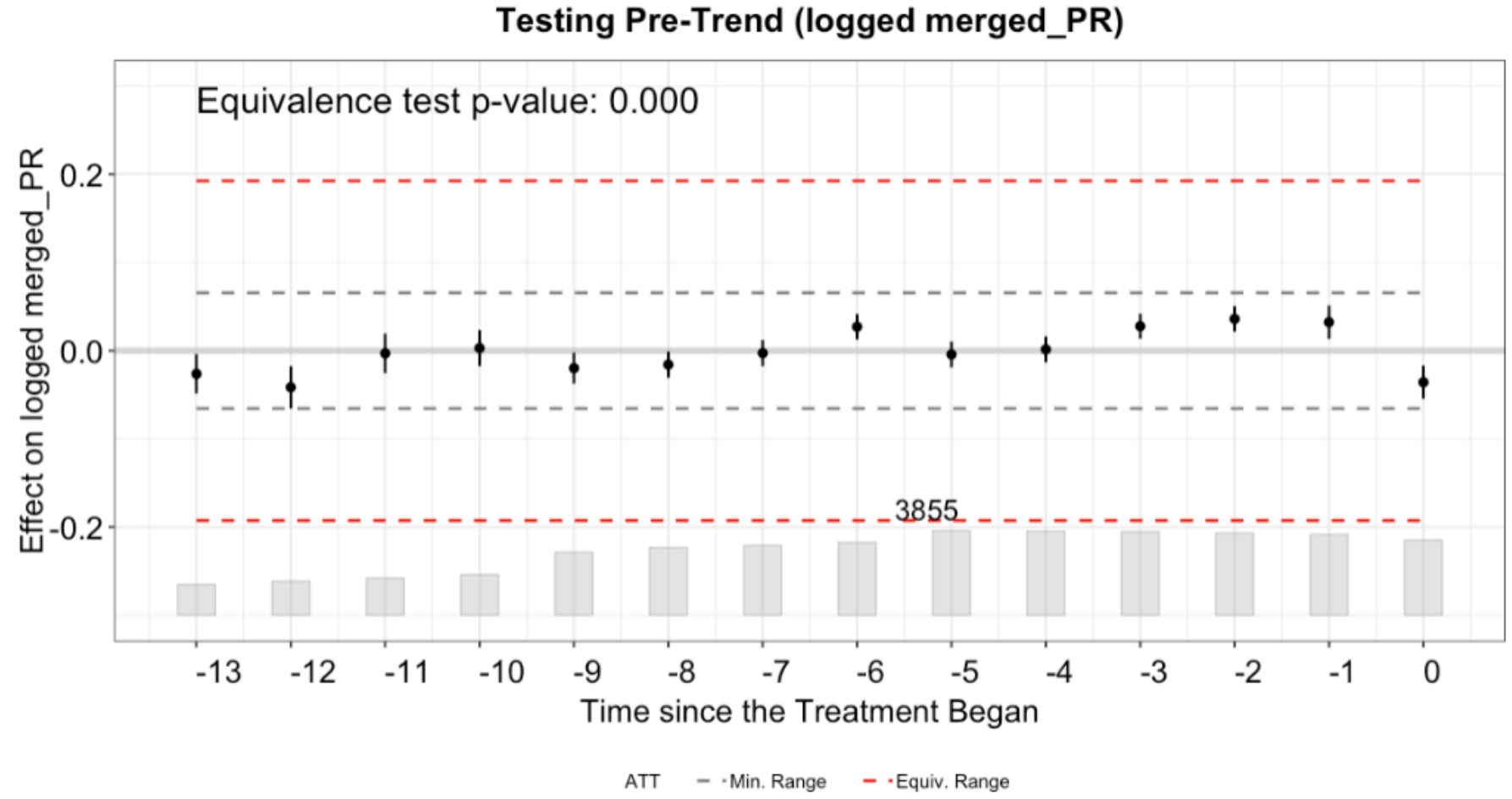

Figure B.1: Pre-trend test (TOST) of logged merged_PR

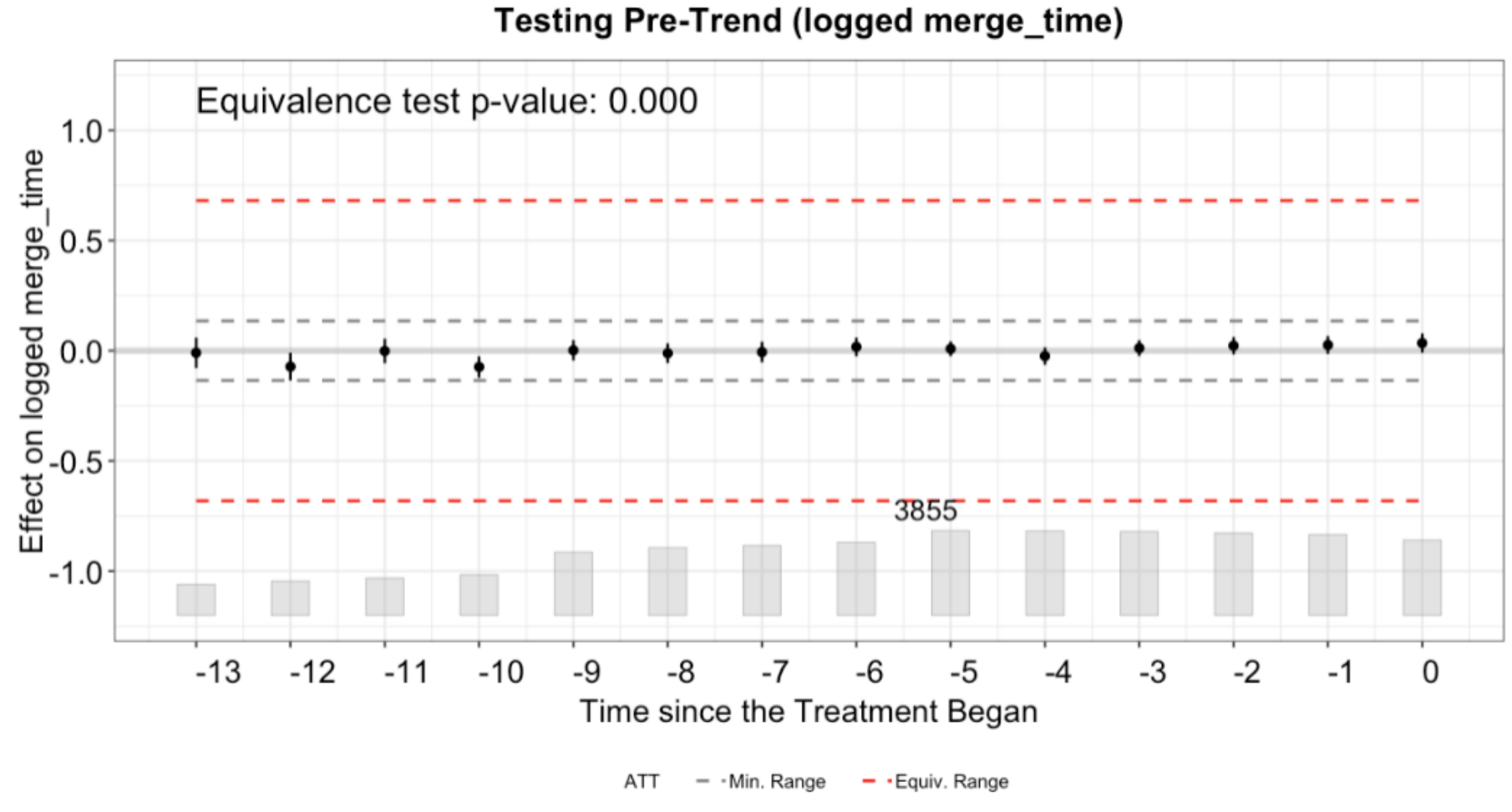

Figure B.2: Pre-trend test (TOST) of logged merge_time

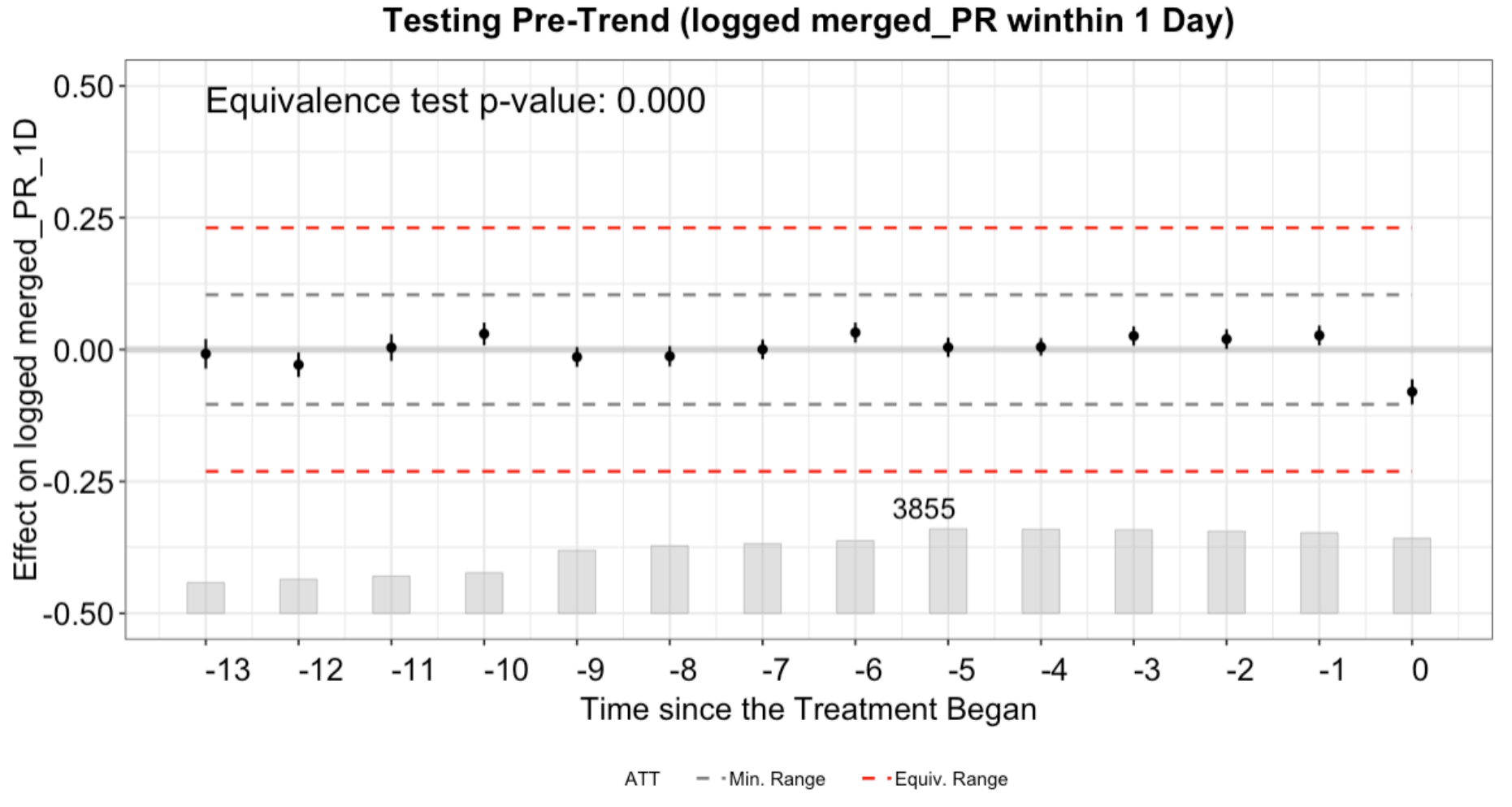

Figure B.3: Pre-trend test (TOST) of logged merged_PR_1D

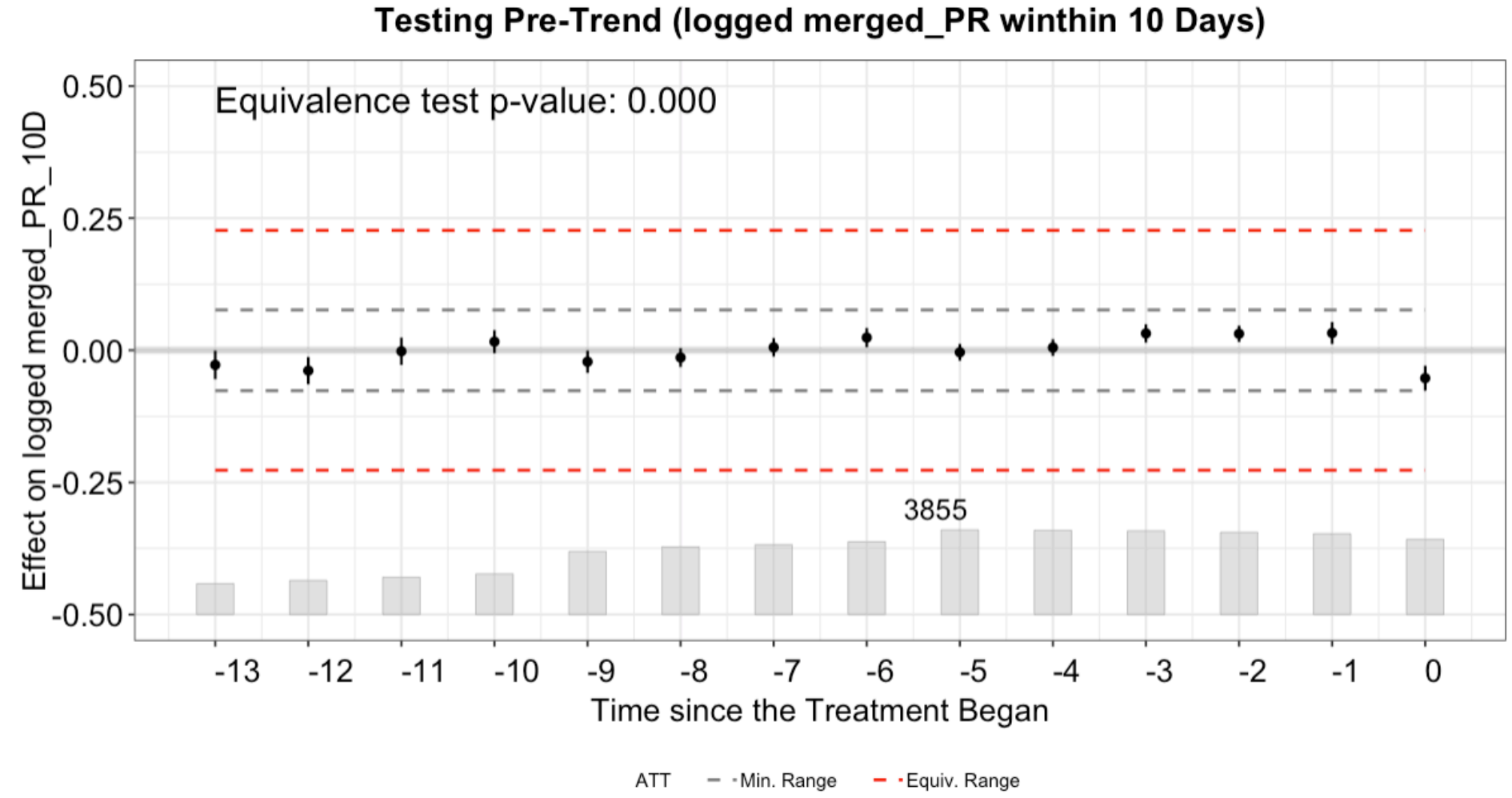

Figure B.4: Pre-trend test (TOST) of logged merged_PR_3D

Figure B.5: Pre-trend test (TOST) of logged merged_PR_10D

**Appendix C: Placebo Test in GSCM**

To further validate our causal identification strategy based on GSCM, we conduct placebo tests by examining pre-treatment periods. Specifically, we define a placebo window covering the three periods prior to treatment and re-estimate the model to assess whether any spurious effects arise before the actual adoption of Copilot. These falsification tests help evaluate the strict exogeneity assumption and detect potential model misspecification. If our identification strategy is valid and the observed effects are driven by Copilot adoption, we should not observe meaningful effects in the pre-treatment period.

To statistically evaluate the results from our placebo analyses, we apply the two-one-sided t (TOST) equivalence test, which assesses whether the estimated placebo effects are practically equivalent to zero (Liu et al. 2024). This approach specifies an equivalence interval around zero and tests the null hypothesis that the placebo effects are meaningfully different from zero.

The results from these tests for the main outcome variables, including project-level code contributions, coordination time and overall project-level timely merge of code contributions, are shown in Figures C.1, through C.5. We observe that the estimated effects in the pre-treatment placebo are at or below zero, while those from the post-treatment period are substantially above zero. The equivalence tests reject the null of inequivalence, indicating that there are no meaningful pre-treatment effects. These findings support the validity and robustness of our causal identification strategy.

It is worth noting that there is a drop at the onset of treatment (i.e., time = 0) in contribution-related measures, such as the number of merged PRs and the number of timely merged PRs. This drop likely reflects an initial adjustment period as developers adapt to the AI pair programmer or integrate it into their workflow. Importantly, the equivalence holds in the pre-treatment window, supporting the credibility of the synthetic control construction.

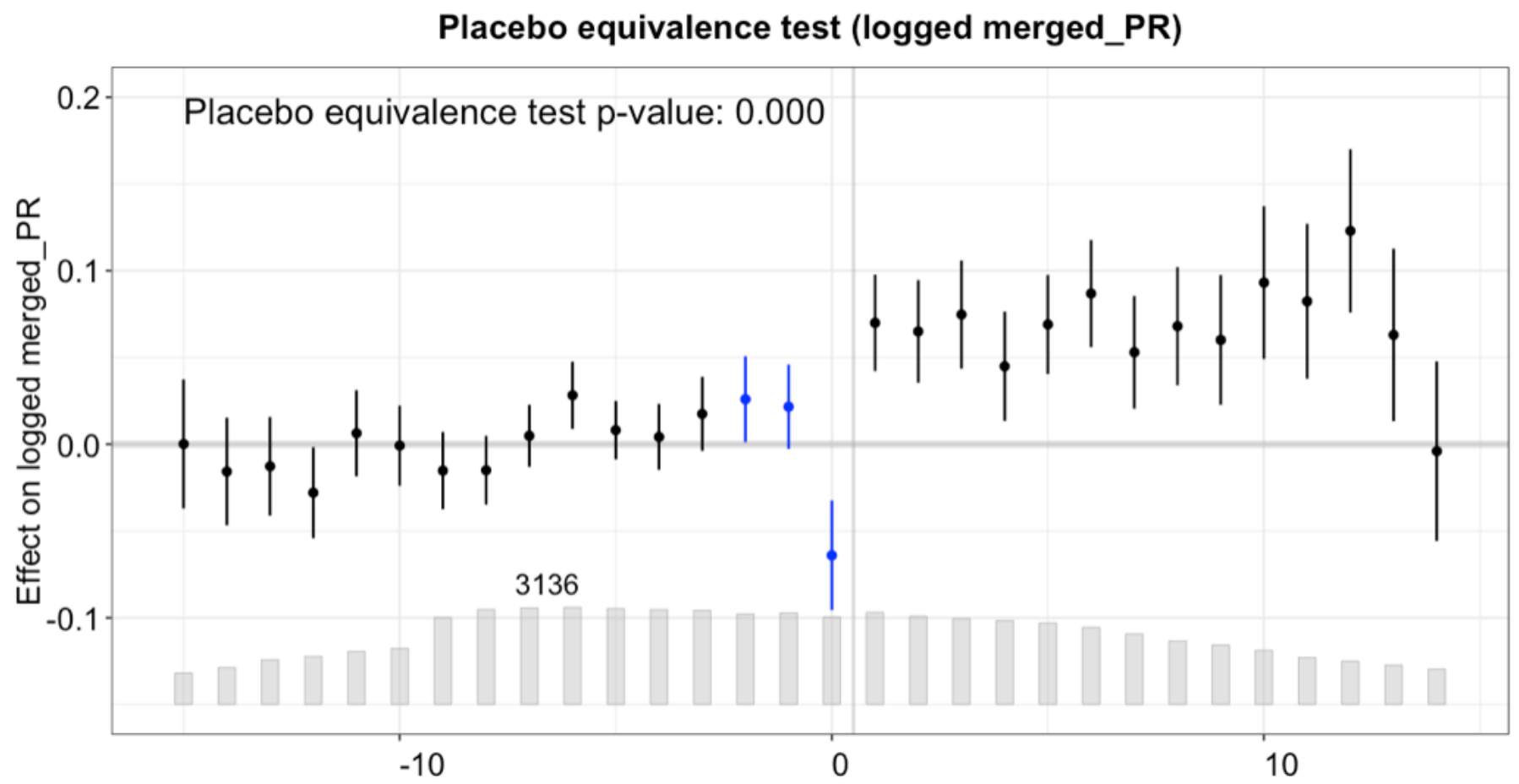

Figure C.1 Placebo Equivalence Test of logged merged_PR

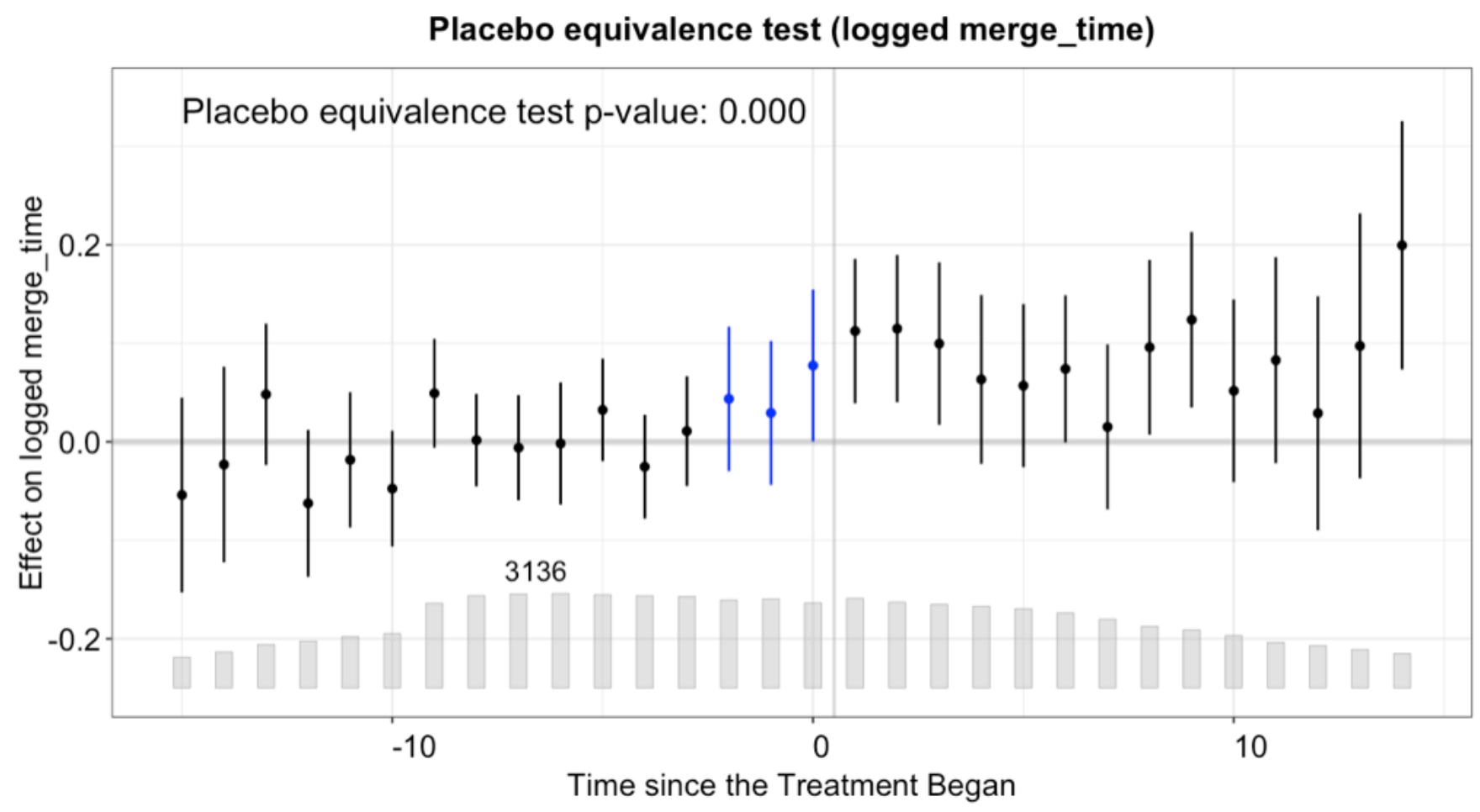

Figure C.2: Placebo Equivalence Test of logged merge_time

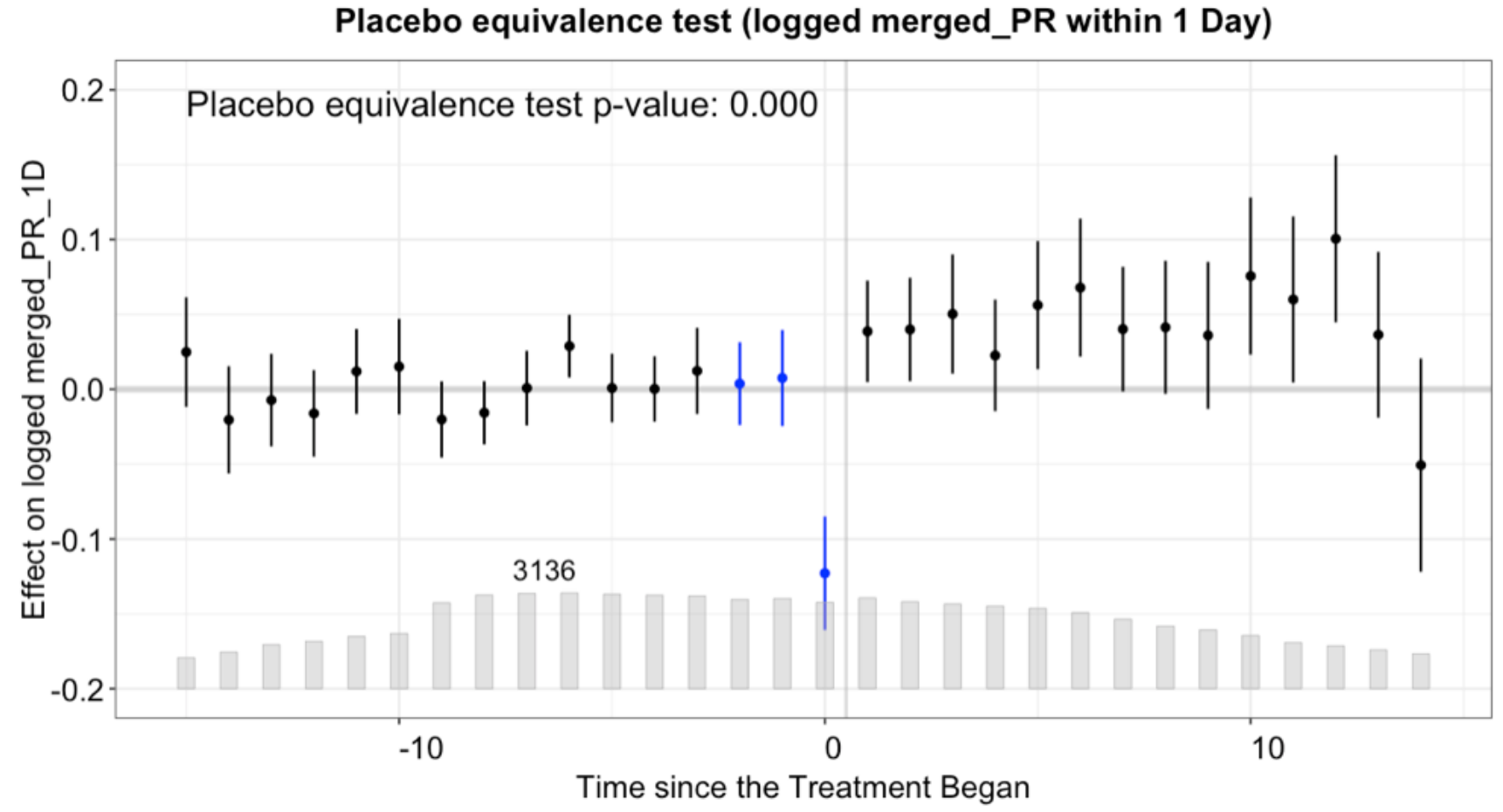

Figure C.3: Placebo Equivalence Test of logged merged_PR_1D

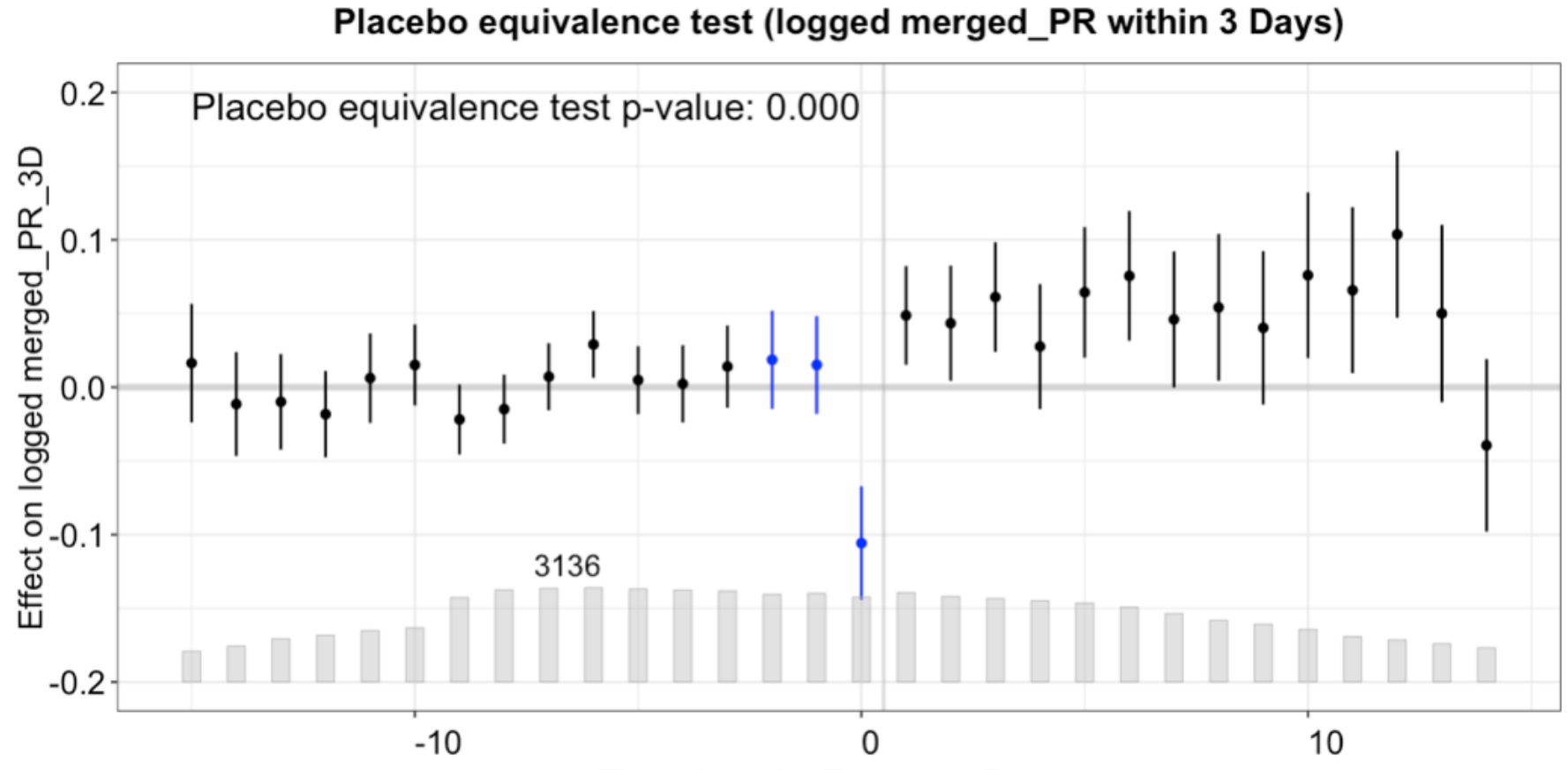

Figure C.4: Placebo Equivalence Test of logged merged_PR_3D

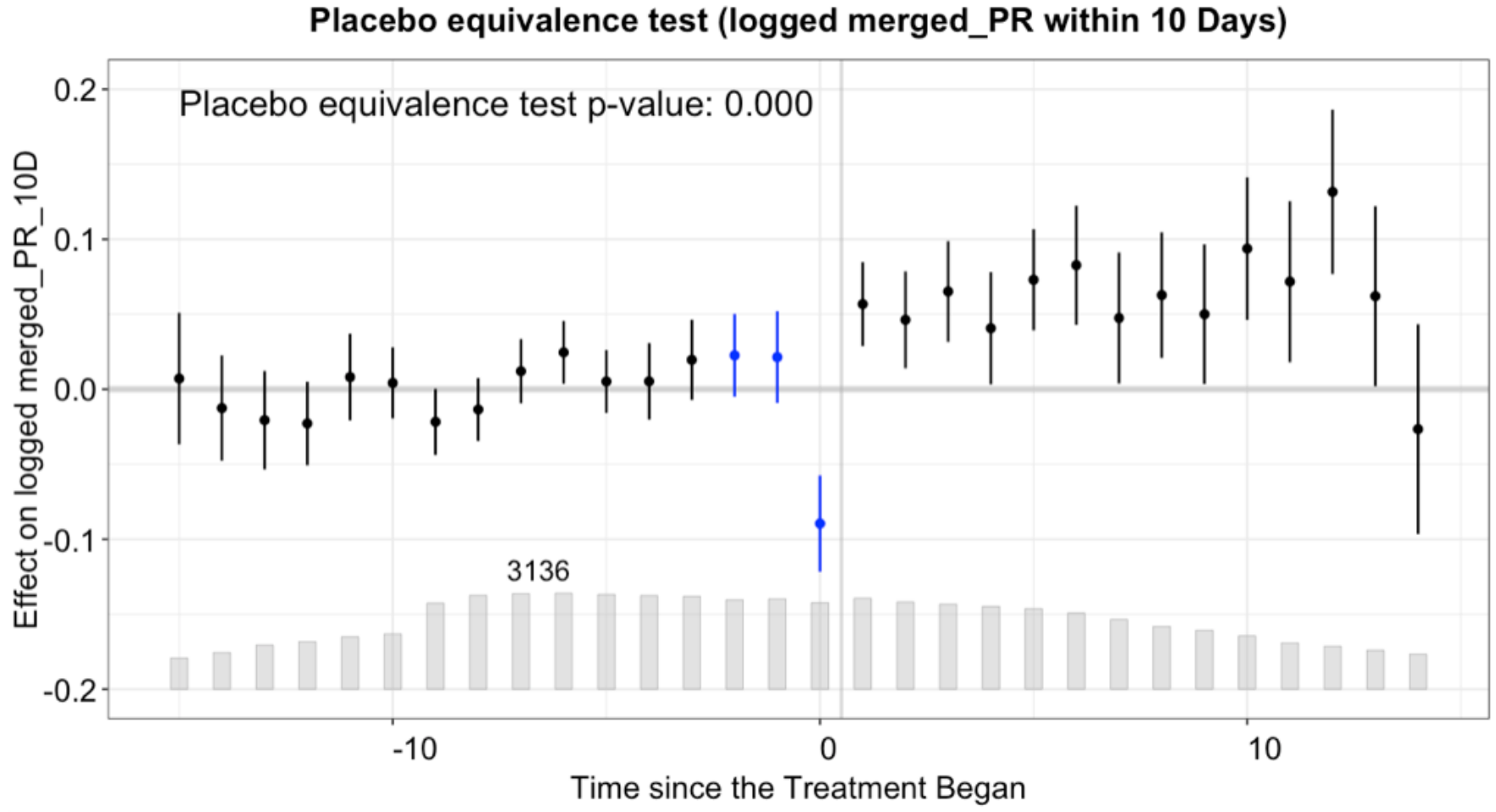

Figure C.5: Placebo Equivalence Test of logged merged_PR_10D

**Appendix D: Within-IDE Analysis**

In our baseline identification, we use IDEs that supported Copilot during our sample period as part of the criteria to identify projects in the treatment group and consider projects developed with IDEs that did not support Copilot for the control group. To mitigate concerns regarding potential unobservable differences in IDE usage, we conduct the analysis by only focusing on projects using Copilot-supported IDEs and further restricting the treatment versus control group comparison within the same Copilot-supported IDE. This approach addresses three key sources of potential bias. First, it controls for the differences between projects using supported versus unsupported IDEs, the latter of which may be associated with lower popularity. Second, it controls for variation across different supported IDEs, which may attract distinct types of projects or developer teams. Third, it accounts for the possibility that developers who choose IDEs compatible with Copilot may differ from those who do not. By focusing on comparing projects developed within the same Copilot-supported IDE, we ensure that both treatment and control groups share the same development environment and that contributing developers are likely to have similar preferences and characteristics. Therefore, the only variation across groups lies in the timing of Copilot adoption.

Specifically, our first step is to identify the sample of projects using Copilot-supported IDEs. Then, in order to identify treatment and control groups, we use projects that adopted Copilot in the first half of the period from July 2021 to December 2022 when Copilot was available, i.e., July 2021 - March 2022, as the treatment group. Accordingly, we conduct this analysis using only the period from January 2021 to March 2022, so projects that adopted Copilot after March 2022 can be considered as the control group. We choose March 2022 as the endpoint so that there are enough projects with Copilot usage during this time window (i.e., a total of nine months from July 2021 to March 2022). Our second step is to match treatment and control projects within the same IDE. After matching, we use the pooled sample to run the GSCM model. This analysis includes 3,215 projects in the treatment group and 1,247 projects in the control group, a total of 4,462 projects. The treatment turn-on time for each project in the treatment group is decided in the same manner as in our baseline analysis. The results based on estimation using GSCM in Table D.1 show that Copilot increased the number of merged PRs and the average time taken to merge a single PR, which are consistent with our main findings.

Table D.1: The Impact of Copilot on Project-Level Code Contributions and Coordination Time (Within-IDE Analysis)

| | (1)<br>Log (Merged_PR) | (2)<br>Log (Merge_time) |
|---|---|---|
| ATT of Copilot | 0.031*** | 0.055** |
| | (0.012) | (0.027) |
| Project FE | Yes | Yes |
| Month FE | Yes | Yes |
| # of projects | 4,462 | 4,462 |
| Observations | 60,659 | 60,659 |

Notes: All estimations are based on the GSCM method. Robust standard errors clustered at project level in parentheses. Please refer to Table 1 for detailed definitions of the dependent variables used in this table. *** p<0.01, ** p<0.05, * p<0.1.

## Appendix E: PSM and DID

We further validate our results using an alternative matching and estimation approach. Specifically, we use the difference-in-differences (DID) estimation combined with the propensity score matching (PSM). In this analysis, we use June 2021, the first availability date of Copilot, as the single treatment turn-on time for all projects classified into the treatment group in the baseline analysis. Hence, this approach may alleviate concerns about non-random treatment turn-on time for Copilot usage in the treatment group in the main analysis. However, we acknowledge that because developers may sign up gradually after the initial availability date, this approach may consider the period when Copilot was actually not used as the post-treatment period, leading to an underestimation of the positive effect.

We calculate each project's propensity score, defined as the probability of adopting Copilot, using a logit regression model. To construct the covariates for matching, we calculate the monthly values of several project characteristics over the pre-treatment period (January to June 2021), average them across the six-month window, and then apply a log transformation. Specifically, we use the log number of merged PRs, the log average time taken to merge a single PR, the log number of unique developers whose PRs have been merged, the log number of push events, the log number of release events, the log number of opened issues, and the log number of closed issues.

To construct a matched sample, we employ a one-to-one nearest neighbor matching approach without replacement, pairing each treated project that adopted Copilot with a single control project that did not use Copilot. Matching without replacement ensures that each control project is used only once, preventing any single control project from disproportionately influencing the results. While this may reduce the total number of matches, it improves the integrity of the comparison by limiting reuse of controls. Following this matching process between projects where Copilot was used and those that Copilot was not used, we arrive at a dataset comprising 2,122 projects, consisting of 1,061 treatments and 1,061 controls. We report the details of the balance checks of our matched sample in Table E.1. Overall, there are no statistically significant differences between the treatment and control groups across these pre-treatment covariates after matching.

Table E.1: Balance Check of Matched Sample

| Variable | Treatment Mean | Control Mean | Difference | P-value |
|---|---|---|---|---|
| Log (Merged_PR) | 2.10 | 2.05 | 0.05 | 0.20 |
| Log (Merge_time) | 4.12 | 4.09 | 0.03 | 0.60 |
| Log (Num_devs_with_mergedPR) | 1.32 | 1.29 | 0.03 | 0.13 |
| Log (Push) | 2.96 | 2.91 | 0.05 | 0.33 |
| Log (Release) | 0.28 | 0.27 | 0.01 | 0.79 |
| Log (Open_issue) | 1.13 | 1.11 | 0.02 | 0.68 |
| Log (Closed_issue) | 1.05 | 1.04 | 0.01 | 0.79 |

Based on the matched sample, we analyze the impact of Copilot on the number of merged PRs and the average time taken to merge a single PR using the following regression specification:

$$Y_{it} = \beta_0 + \beta_1 Copilot_i \times Post_t + \alpha_i + \mu_t + \varepsilon_{it} \quad (5)$$

where the dependent variable $Y_{it}$ represents the log number of merged PRs and the log average time taken to merge a PR of project $i$ in month $t$. $Copilot_i$ is a binary indicator that is denoted with "1" for project $i$ where Copilot was used and "0" otherwise. $Post_t$ is a binary indicator that is set to "1" in months after June 2021 and "0" otherwise. To control for time-invariant heterogeneity across projects and time trends, we include project-level fixed-effects $\alpha_i$ and month-level fixed-effects $\mu_t$.

The main coefficient of interest is $\beta_1$, as it indicates the change in the number of merged PRs or the change in the average time taken to merge a PR of projects before and after the availability of Copilot, relative to the changes of projects where Copilot was not used at all throughout the sample period. We report the results in Table E.2. The results show that Copilot significantly increased the number of merged PRs and the average time taken to merge a single PR, which are qualitatively consistent with our findings from the main analysis.

Table E.2: The Impact of Copilot on Project-Level Code Contributions and Coordination Time (PSM and DID)

| | (1)<br>Log (Merged_PR) | (2)<br>Log (Merge_time) |
|---|---|---|
| ATT of Copilot | 0.172*** | 0.148*** |
| | (0.027) | (0.043) |
| Project FE | Yes | Yes |
| Month FE | Yes | Yes |
| # of projects | 2,122 | 2,122 |
| Observations | 42,251 | 42,251 |

Notes: All estimations are based on the GSCM method. Robust standard errors clustered at project level in parentheses. Please refer to Table 1 for detailed definitions of the dependent variables used in this table. *** p<0.01, ** p<0.05, * p<0.1.

We next examine whether there is any pre-treatment trend in the analysis of PSM and DID, i.e., whether the outcome variables of projects in the treatment group are different from that in the control group even before the treatment (i.e., the availability of Copilot) happens. The Relative Time Model (RTM) has been used in the economics and information systems literature to validate the pre-treatment parallel trend assumption (Lu et al. 2019, Alyakoob and Rahman 2022). More specifically, we use the following RTM:

$$Y_{it} = \beta_0 + \sum_{\eta=1}^{q} \beta_{1,\eta} \left(Copilot_i \times Post_{t-\eta}\right) + \sum_{\eta=0}^{p} \beta_{2,\eta} \left(Copilot_i \times Post_{t+\eta}\right) + \alpha_i + \mu_t + \varepsilon_{it} \quad (6)$$

In this model, the sum on the right-hand side represents $q$ lead indicators (anticipatory effects) and $p$ lagged indicators (posttreatment effects) where $Y_{it}$ denotes the log number of merged PRs or the log average time taken to merge a single PR for project $i$ in month $t$. For our analysis of PSM and DID, we set $q = 4$ and $p = 18$. When the pre-treatment trend exists, the lead indicators would be able to predict changes in dependent variables, the log number of merged PRs and the log average time taken to merge a PR. Conversely, if the pre-treatment trend is unlikely to be a concern, the lead indicators would not predict dependent variables (Autor 2003, Angrist and Pischke 2009). The estimation results of the RTM are shown in Table E.3. We observe that the coefficients for the four lead indicators are statistically insignificant, indicating that there does not exist a pre-treatment trend.

Table E.3: Pre-trend Test for PSM and DID Analysis

| | (1)<br>PSM and DID<br>Log (Merged_PR) | (2)<br>PSM and DID<br>Log (Merge_time) |
|---|---|---|
| $Copilot_i \times Pre4m$ | -0.033 | 0.037 |
| | (0.034) | (0.088) |
| $Copilot_i \times Pre3m$ | -0.020 | 0.094 |
| | (0.038) | (0.090) |
| $Copilot_i \times Pre2m$ | -0.050 | 0.089 |
| | (0.041) | (0.094) |
| $Copilot_i \times Pre1m$ | -0.031 | 0.100 |
| | (0.041) | (0.094) |
| $Copilot_i \times Post0m$ | 0.036 | 0.165* |
| | (0.044) | (0.098) |
| $Copilot_i \times Post1m$ | 0.068 | 0.100 |
| | (0.046) | (0.101) |
| $Copilot_i \times Post2m$ | 0.097** | 0.090 |
| | (0.047) | (0.100) |
| $Copilot_i \times Post3m$ | 0.042 | 0.237** |
| | (0.046) | (0.116) |
| $Copilot_i \times Post4m$ | 0.159*** | 0.325*** |
| | (0.047) | (0.100) |
| $Copilot_i \times Post5m+$ | 0.175*** | 0.257*** |
| | (0.042) | (0.075) |

Notes: Robust standard errors clustered at project level in parentheses. *** p<0.01, ** p<0.05, * p<0.1.

**Appendix F: Instrumental Variable Analysis**

To address potential endogeneity concerns on Copilot use, we conduct an instrumental variable estimation. Our instrument exploits variation in developers' exposure to Copilot through their past collaborators outside the focal project. The instrument is constructed by first identifying the historical collaborators of each developer in the focal project based on their collaboration activities in external projects. We then combine these developer level collaborator networks to obtain the set of prior collaborators associated with the focal project. Next, we identify whether any of these prior collaborators were early Copilot adopters. If the focal project is connected to one or more early adopters, we calculate the length of time between the earliest collaborator's Copilot adoption and the focal project's own adoption. Following Sharma et al. (2025), these two components, namely exposure to an early adopter and the corresponding exposure window, are combined into a single project-month level instrument used in Table F.1, denoted as *Early Copilot Adopter Exposure*.

We believe the first component satisfies the relevance condition because prior research shows that individuals learn from peers through social interactions and shared experience (Foster and Rosenzweig 1995, Conley and Udry 2010, Bollinger and Gillingham 2012, Bailey et al. 2022). Learning about new technologies often occurs through observing others' outcomes and updating beliefs about their profitability, especially under uncertainty, and such learning can diffuse through social and professional networks (Foster and Rosenzweig 1995, Conley and Udry 2010). In addition, social interactions can generate spillovers in adoption decisions, where exposure to peers who have adopted a new technology increases the likelihood of one's own adoption (Bollinger and Gillingham 2012, Bailey et al. 2022). Therefore, developers who had past collaborators who were early Copilot users are more likely to be exposed to relevant knowledge and experiences, increasing their likelihood of adoption.

The second component satisfies the relevance condition by capturing the temporal accumulation of social learning. Prior work suggests that the benefits of new technologies grow as knowledge accumulates over time (Foster and Rosenzweig 1995, Conley and Udry 2010). Moreover, peer effects are persistent and can unfold over extended periods, as individuals continue to be influenced by exposure to others' adoption behaviors (Bollinger and Gillingham 2012, Bailey et al. 2022). Consistent with this evidence, a longer gap between a collaborator's early adoption and the focal project's adoption provides more opportunities for information exchange and observation, thereby increasing the likelihood of Copilot adoption.

Moreover, this instrumental variable meets the exclusion restriction. Early adoption occurs in unrelated projects outside our analytical sample and thus does not directly affect outcomes in the focal project. Instead, its influence operates only through the focal developers' adoption decision. Because the instrument is constructed from historical collaborations external to the focal project, it is unlikely to be correlated with project level outcomes such as merged PRs or average time to merge a single PR.

Specifically, based on our sample, we define early adopters as developers who adopted Copilot between July and September 2021, shortly after its initial release in June 2021. For a focal developer in a focal project between October 2021 and December 2022, we identify their prior collaborators based on the collaboration activities in external projects in 2020. If any of these external collaborators adopted Copilot during the early period (i.e. July to September 2021), the focal developer is considered exposed to early adoption. We then measure the time between the external collaborator's adoption and the focal project's adoption as the exposure window.

We report the results in Table F.1. The first stage shows a significantly positive relationship between the instrumental variable and Copilot use, with an F-statistic of 268.36, indicating strong instrument relevance. The second stage results indicate that Copilot increased both the number of merged PRs and the average time to merge a single PR, consistent with our main findings. Nevertheless, we acknowledge that developers connected to early adopters may differ in other ways, such as network characteristics or exposure to other technologies. Accordingly, we view this instrumental variable analysis as supplementary to our main identification strategy.

Table F.1: The Impact of Copilot on Project-Level Code Contributions and Coordination Time (Instrumental Variable Analysis)

| First-stage: | (1)<br>Copilot Use | Second-stage: | (2)<br>Log (merged_PR) | (3)<br>Log (merge_time) |
|---|---|---|---|---|
| Early Copilot Adopter Exposure | 0.091***<br>(0.006) | ATT of Copilot | 0.070***<br>(0.017) | 0.086**<br>(0.038) |
| F-statistics | 268.36 | | | |
| Project FE | Yes | Project FE | Yes | Yes |
| Month FE | Yes | Month FE | Yes | Yes |
| # of projects | 2,493 | # of projects | 2,493 | 2,493 |
| Observations | 37,380 | Observations | 37,380 | 37,380 |

Robust standard errors clustered at project-month level in parentheses. *** p<0.01, ** p<0.05, * p<0.1.

## Appendix G: Additional Robustness Checks

### G.1 Refined Sample

As developers work on multiple projects, it is possible that some may be working on both treatment and control projects. Approximately 3.5% of the selected projects in our main analysis have developers involved in both groups. Although these developers might not be using Copilot for the control projects due to IDE restrictions, there is potential for knowledge transfer, which could bias our results. To eliminate this bias, we refine our sample to exclusively assess the impact of Copilot on projects where no developer is involved in both groups. As shown in Tables G.1, the results indicate that Copilot increased the number of merged PRs and the average time taken to merge a single PR, which are qualitatively similar to our main results.

There may be spillover effects wherein projects not using Copilot could still benefit if their developers have prior experience with Copilot from other projects. In such cases, our estimates would represent a lower bound on the true impact of Copilot on code contributions and coordination time. The observed increases in code contributions and coordination time would be more pronounced when compared to projects unaffected by such spillover effects.

Table G.1: The Impact of Copilot on Project-Level Code Contributions and Coordination Time (Refined Sample)

| | (1)<br>Log (Merged_PR) | (2)<br>Log (Merge_time) |
|---|---|---|
| ATT of Copilot | 0.062*** | 0.080*** |
| | (0.007) | (0.022) |
| Project FE | Yes | Yes |
| Month FE | Yes | Yes |
| # of projects | 7,369 | 7,369 |
| Observations | 135,731 | 135,731 |

Notes: All estimations are based on the GSCM method. Robust standard errors clustered at project level in parentheses. Please refer to Table 1 for detailed definitions of the dependent variables used in this table. *** p<0.01, ** p<0.05, * p<0.1.

### G.2 Outlier Removal

To rule out the possibility that extreme values are disproportionately influencing the results, we conduct an analysis that excludes projects exhibiting unusually high or low values in terms of the number of merged PRs and the average time taken to merge a single PR. By removing these outliers, we ensure that our results reflect generalizable trends rather than being driven by a few atypical cases. The results, reported in Table G.2, remain consistent with our main findings.

Table G.2: The Impact of Copilot on Project-Level Code Contributions and Coordination Time (Outlier Removal)

| | (1)<br>Log (Merged_PR) | (2)<br>Log (Merge_time) |
|---|---|---|
| ATT of Copilot | 0.060*** | 0.074*** |
| | (0.008) | (0.022) |
| Project FE | Yes | Yes |
| Month FE | Yes | Yes |
| # of projects | 7,557 | 7,557 |
| Observations | 138,123 | 138,123 |

Notes: All estimations are based on the GSCM method. Robust standard errors clustered at project level in parentheses. Please refer to Table 1 for detailed definitions of the dependent variables used in this table. *** p<0.01, ** p<0.05, * p<0.1.

### G.3 Non-AI Topic Projects

Another potential concern is that the observed effects may be driven by project-specific enthusiasm associated with certain topics, particularly those related to artificial intelligence. AI-related projects may attract different groups of developers compared to projects focused on other topics, which could confound the estimated effect of Copilot adoption. To address this concern, we conduct a robustness check by excluding projects associated with AI-related topics and re-estimate the models using the remaining sample. This approach allows us to test whether the impact of Copilot is generalizable across a broader range of project topics and not limited to those inherently aligned with AI development. As shown in Table G.3, the results remain consistent with our main findings.

Table G.3: The Impact of Copilot on Project-Level Code Contributions and Coordination Time (Non-AI Topic Projects)

| | (1)<br>Log (Merged_PR) | (2)<br>Log (Merge_time) |
|---|---|---|
| ATT of Copilot | 0.057*** | 0.088*** |
| | (0.008) | (0.021) |
| Project FE | Yes | Yes |
| Month FE | Yes | Yes |
| # of projects | 6,668 | 6,668 |
| Observations | 121,366 | 121,366 |

Notes: All estimations are based on the GSCM method. Robust standard errors clustered at project level in parentheses. Please refer to Table 1 for detailed definitions of the dependent variables used in this table. *** p<0.01, ** p<0.05, * p<0.1.

### G.4 Alternative Measures

In the main analysis, we measure project-level code contributions using merged PRs and coordination time using hours between PR submission and acceptance. To ensure our results are not driven by specific measures, we examine alternative measures, including submitted PRs and commits for project-level code contributions, and time intervals in days and minutes for coordination time. Results in Tables G.4.1 and G.4.2 remain consistent, indicating our findings are robust to outcome variable definitions.

Table G.4.1: The Impact of Copilot on Project-Level Code Contributions

(Alternative Measures of Code Contributions)

| | (1) GSCM Main Sample Log (Submitted PRs) | (2) GSCM Main Sample Log (Commits) |
|---|---|---|
| ATT of Copilot | 0.075*** | 0.107*** |
| | (0.009) | (0.014) |
| Project FE | Yes | Yes |
| Month FE | Yes | Yes |
| # of projects | 7,637 | 7,637 |
| Observations | 139,329 | 139,329 |

Notes: All estimations are based on the GSCM method. Robust standard errors clustered at project level in parentheses. *** p<0.01, ** p<0.05, * p<0.1.

Table G.4.2: The Impact of Copilot on Project-Level Coordination Time

(Alternative Measures of Coordination Time)

| | (1) GSCM Main Sample Log (Merge time in days) | (2) GSCM Main Sample Log (Merge time in minutes) |
|---|---|---|
| ATT of Copilot | 0.049*** | 0.101*** |
| | (0.016) | (0.032) |
| Project FE | Yes | Yes |
| Month FE | Yes | Yes |
| # of projects | 7,637 | 7,637 |
| Observations | 139,329 | 139,329 |

Notes: All estimations are based on the GSCM method. Robust standard errors clustered at project level in parentheses. *** p<0.01, ** p<0.05, * p<0.1.

### G.5 Effect of Workload

One important alternative explanation for the observed increase in coordination time for merging a PR following Copilot adoption is that there were simply more PRs that needed to be reviewed, which could lead to some delay. To address this concern, we include the log number of cumulative submitted PRs till the focal month as a control variable to account for potential increases in overall workload when analyzing coordination time. The results, reported in Tables G.5, are qualitatively similar to our main results for coordination time.

Table G.5: The Impact of Copilot on Project-Level Coordination Time with Code Submission

| | (1) GSCM Main Analysis Log (Merge_time) | (2) GSCM Within-IDE Analysis Log (Merge_time) | (3) GSCM Refined Sample Log (Merge_time) | (4) GSCM Outlier Removal Log (Merge_time) | (5) GSCM Non-AI Topic Log (Merge_time) | (6) PSM and DID Single Treatment Turn-on Time Log (Merge_time) |
|---|---|---|---|---|---|---|
| ATT of Copilot | 0.079*** | 0.055** | 0.077*** | 0.072*** | 0.087*** | 0.147*** |
| | (0.023) | (0.028) | (0.021) | (0.023) | (0.021) | (0.044) |
| Log (Open_PR) | 0.031 | 0.041 | 0.054* | 0.021 | 0.028 | 0.057 |
| | (0.029) | (0.041) | (0.031) | (0.032) | (0.032) | (0.037) |
| Project FE | Yes | Yes | Yes | Yes | Yes | Yes |
| Month FE | Yes | Yes | Yes | Yes | Yes | Yes |
| # of projects | 7,637 | 4,462 | 7,369 | 7,557 | 6,668 | 2,122 |
| Observations | 139,329 | 60,659 | 135,731 | 138,123 | 121,366 | 42,251 |

Notes: *Open_PR* refers to the total number of cumulative submitted PRs till the focal month and is added to the regressions to control for the effect of workload. Robust standard errors clustered at project level in parentheses. *** p<0.01, ** p<0.05, * p<0.1.

### G.6 Alternative Estimation

We also validate our results using a two-way fixed effects model applied to the full sample. The findings presented in Table G.6 are qualitatively consistent with our main analysis.

Table G.6: The Impact of Copilot on Project-Level Code Contributions and Coordination Time (Two-way Fixed Effects with Full Sample)

| | (1) TWFE Project-level Code Contributions | (2) TWFE Coordination Time |
|---|---|---|
| | Log (Merged_PR) | Log (Merge_time) |
| ATT of Copilot | 0.029*** | 0.036** |
| | (0.008) | (0.018) |
| Project FE | Yes | Yes |
| Month FE | Yes | Yes |
| # of projects | 9,244 | 9,244 |
| Observations | 140,084 | 140,084 |

Notes: Robust standard errors clustered at project level in parentheses. *** p<0.01, ** p<0.05, * p<0.1.

### G.7 Expanded Sample

Because our main sample includes projects with three or more developers, we also examine the robustness of our results using an expanded sample that includes projects with two developers. Results are reported in Table G.7 and are qualitatively similar to our main results.

Table G.7: The Impact of Copilot on Project-Level Code Contributions and Coordination Time
(Expanded Sample that Includes Projects with Two Developers)

| | (1)<br>GSCM Expanded Sample<br>Project-level Code Contributions | (2)<br>GSCM Expanded Sample<br>Coordination Time |
|---|---|---|
| | Log (Merged_PR) | Log (Merge_time) |
| ATT of Copilot | 0.013** | 0.083*** |
| | (0.006) | (0.015) |
| Project FE | Yes | Yes |
| Month FE | Yes | Yes |
| # of projects | 12,512 | 12,512 |
| Observations | 240,433 | 240,433 |

Notes: All estimations are based on the GSCM method. Robust standard errors clustered at project level in parentheses. *** p<0.01, ** p<0.05, * p<0.1.

### G.8 Additional Alternative Sample

In our main analysis, we restrict the sample to project-month observations in which at least one PR was eventually merged. To assess the robustness of our findings, we replicate the analysis by including all project-month observations, regardless of whether any PRs were merged. As shown in Table G.8, the results remain consistent with our main findings. However, it is important to note that we cannot extend this analysis to coordination time as measuring coordination time requires data on PR acceptance time, which is only available when PR is merged.

Table G.8: The Impact of Copilot on Project-Level Code Contributions
(Expanded Sample that Includes Project-month Observations with Zero Merged PR)

| | Log (Merged_PR) |
|---|---|
| ATT of Copilot | 0.066*** |
| | (0.006) |
| Project FE | Yes |
| Month FE | Yes |
| # of projects | 8,965 |
| Observations | 206,897 |

Notes: All estimations are based on the GSCM method. Robust standard errors clustered at project level in parentheses. *** p<0.01, ** p<0.05, * p<0.1.

**Appendix H: Additional Evidence on the Mechanisms for H1 and H2**

Table H.1: How Developer Coding Participation and Individual Productivity are Correlated with Project-Level Code Contributions

| | (1)<br>Log (Merged_PR) | (2)<br>Log (Merged_PR) |
|---|---|---|
| Log (Num_devs_with_mergedPR) | 1.213***<br>(0.006) | |
| Log (MergedPR_per_dev) | | 1.173***<br>(0.006) |
| Project FE | Yes | Yes |
| Month FE | Yes | Yes |
| # of projects | 7,637 | 7,637 |
| Observations | 139,329 | 139,329 |

Notes: Robust standard errors clustered at project level in parentheses. *** p<0.01, ** p<0.05, * p<0.1.

Table H.2: How Code Discussion is Correlated with Project-Level Coordination Time

| | (1)<br>Log (Merge_time) | (2)<br>Log (Merge_time) | (3)<br>Log (Merge_time) |
|---|---|---|---|
| Log (Comment_per_mergedPR) | 0.059***<br>(0.004) | | |
| Log (Num_devs_with_comments_per_mergedPR) | | 0.259***<br>(0.016) | |
| Log (Comments_per_dev_per_mergedPR) | | | 0.063***<br>(0.004) |
| Project FE | Yes | Yes | Yes |
| Month FE | Yes | Yes | Yes |
| # of projects | 7,637 | 7,637 | 7,637 |
| Observations | 139,329 | 139,329 | 139,329 |

Notes: Robust standard errors clustered at project level in parentheses. *** p<0.01, ** p<0.05, * p<0.1.

Table H.3: How Project-Level Code Contributions and Coordination Time are Correlated with Project-Level Timely Merge of Code Contributions

| | (1)<br>Log (MergedPR _1D) | (2)<br>Log (MergedPR _1D) | (3)<br>Log (MergedPR _3D) | (4)<br>Log (MergedPR _3D) | (5)<br>Log (MergedPR _10D) | (6)<br>Log (MergedPR _10D) |
|---|---|---|---|---|---|---|
| Log (Merge_time) | -0.152***<br>(0.002) | | -0.144***<br>(0.002) | | -0.119***<br>(0.002) | |
| Log (Merged_PR) | | 1.027***<br>(0.003) | | 1.058***<br>(0.002) | | 1.069***<br>(0.002) |
| Project FE | Yes | Yes | Yes | Yes | Yes | Yes |
| Month FE | Yes | Yes | Yes | Yes | Yes | Yes |
| # of projects | 7,637 | 7,637 | 7,637 | 7,637 | 7,637 | 7,637 |
| Observations | 139,329 | 139,329 | 139,329 | 139,329 | 139,329 | 139,329 |

Notes: Robust standard errors clustered at project level in parentheses. *** p<0.01, ** p<0.05, * p<0.1.

## Appendix I: Additional Analyses on Code Discussions among Developers

### I.1 Text Analysis of Comment Content

Prior research in team coordination emphasizes that the content of discussions also plays a role in shaping coordination time, as greater diversity in discussion topics can increase the complexity and difficulty of coordination (Hansen et al. 2018). Building on this insight, we argue that the same logic may apply to software development, where developer code discussions serve as a primary mechanism for coordination. In this context, a broader range of topics discussed may introduce additional coordination challenges.

To examine the effect of GitHub Copilot on the diversity of discussion topics within merged PRs, we apply a Latent Dirichlet Allocation (LDA) topic modeling approach. Specifically, we measure both the number of topics discussed per merged PR and the entropy of topic distribution. We adopt the LDA model for three key reasons. First, LDA addresses the data sparsity challenges inherent in traditional approaches such as TF-IDF. Second, the topics and associated keywords generated by LDA are human-interpretable, in contrast to the latent dimensions produced by models like doc2vec. Third, LDA has been widely validated and adopted in prior literature across various research contexts, including scientific publications (Griffiths and Steyvers 2004, Wang and Blei 2011), social media content (Ramage et al. 2010, Weng et al. 2010, Lee et al. 2016, Shin et al. 2020), business descriptions (Shi et al. 2016), and mobile App description (Lee et al. 2020). The key advantage of using LDA lies in its ability to identify latent themes from the data by grouping semantically related keywords into coherent topics, rather than treating each word as an independent feature. We exclude comments that are not in English or are too short to convey meaningful information. After this step, we estimate the topic model using a corpus of 5,404,735 merged PR comments collected from selected projects between 2021 and 2022. This modeling approach provides a multinomial distribution over the inferred topics for each merged PR, enabling us to assess both the number of distinct topics present and the evenness of their distribution.

A critical hyperparameter in LDA modeling is the number of topics to extract. Several methods have been proposed to determine the optimal number, including metrics such as perplexity, topic coherence, and topic diversity. Given our objective of categorizing advice-related content into distinct yet coherent themes, we rely on both coherence and diversity scores to guide model selection. Based on these criteria, we identify 10 as the optimal number of topics, balancing topic distinctiveness and coherence.

The results, presented in Table I.1, indicate that following the adoption of Copilot use, there was an increase in both the number of discussion topics and the entropy score per merged PR. This suggests that Copilot introduces diverse code discussion content for specific changes in each merged PR, leading to increased coordination time.

Table I.1: The Impact of Copilot on Project-Level Code Discussion Content Per Merged PR

| | (1)<br>Entropy_score_per_mergedPR | (2)<br>Topics_count_per_mergedPR |
|---|---|---|
| ATT of Copilot | 0.017** | 0.029** |
| | (0.007) | (0.014) |
| Project FE | Yes | Yes |
| Month FE | Yes | Yes |
| # of projects | 4,009 | 4,009 |
| Observations | 78,636 | 78,636 |

Notes: All estimations are based on the GSCM method. Robust standard errors clustered at project level in parentheses. *** p<0.01, ** p<0.05, * p<0.1.

### I.2 Discussion Cycles

In addition to discussion content, we examine the structure of interactions by analyzing the number of discussion cycles per merged PR, defined as the number of iterative rounds of discussion concerning the same proposed changes within a single merged PR. This measure captures the extent to which a merged PR requires repeated iterations before reaching convergence. The results, reported in Table I.2, indicate that Copilot increased the number of discussion cycles per merged PR. Such iterative adjustments extend the time required to reach agreement and finalize integration, thereby increasing overall coordination time.

Table I.2: The Effect of Copilot on Discussion Cycles per Merged PR

| | (1)<br>Log (discussion_cycles_per_mergedPR) |
|---|---|
| ATT of Copilot | 0.011** |
| | (0.005) |
| Project FE | Yes |
| Month FE | Yes |
| # of projects | 7,637 |
| Observations | 139,329 |

Notes: All estimations are based on the GSCM method. Robust standard errors clustered at project level in parentheses. *** p<0.01, ** p<0.05, * p<0.1.

## Appendix J: The Impact of Copilot and Project Complexity

In OSS communities, complex projects typically contain more features and code modules. Prior work suggests that developers are more likely to contribute to complex projects, as they offer greater status and visibility benefits (Chengalur-Smith et al. 2010, Hann et al. 2013, Medappa and Srivastava 2019). However, these projects also tend to impose greater demands for knowledge sharing, communication, and mutual understanding (Williams 2011, Zimmermann et al. 2018). While generative AI tools enhance developers' capacity to produce code, they do not reduce the inherent complexity of a project. There are two plausible, yet opposing, mechanisms through which project complexity might influence the observed effects of Copilot.

On one hand, complex projects may attract more contributors due to their higher status potential, which can lead to greater project-level code contributions. However, the increased number of contributors also brings a wider range of perspectives and development styles, making it more difficult to reach consensus and resolve conflicting views, thereby increasing coordination time. On the other hand, these projects may present higher expected costs, which could limit participation in both coding and non-coding activities (i.e., code discussion) and reduce coordination time. In either case, it is important to determine whether the effects we observe are simply artifacts of project complexity rather than the result of Copilot adoption. To address this concern, we conduct the analysis by splitting the sample into complex versus simple projects based on the number of lines of code in each project prior to Copilot adoption. We use a median split, classifying projects above the median as complex and those below the median as simple.

The results in Tables J.1 and J.2 show that both simple and complex projects exhibit similar patterns. Copilot adoption is associated with increased project-level code contributions and longer coordination time in both cases. To further examine this, we conduct PSM and DID analyses using an interaction term for project complexity. We define a binary variable "Complexity" that equals 1 if a project's lines of code exceed the median value before the Copilot use and 0 otherwise. As shown in Table J.3, the interaction term is not statistically significant, indicating no difference in Copilot's impact between simple and complex projects.

Table J.1: The Impact of Copilot on Simple Projects

| | (1)<br>Project-level Code Contributions | (2)<br>Coordination Time | (3)<br>Timely Project Outputs (1 Day) | (4)<br>Timely Project Outputs (3 Days) | (5)<br>Timely Project Outputs (10 Days) |
|---|---|---|---|---|---|
| | Log (Merged_PR) | Log (Merge_time) | Log (MergedPR_1D) | Log (MergedPR_3D) | Log (MergedPR_10D) |
| ATT of Copilot | 0.046*** | 0.082** | 0.025** | 0.029*** | 0.039*** |
| | (0.011) | (0.032) | (0.012) | (0.011) | (0.011) |
| Project FE | Yes | Yes | Yes | Yes | Yes |
| Month FE | Yes | Yes | Yes | Yes | Yes |
| # of projects | 3,591 | 3,591 | 3,591 | 3,591 | 3,591 |
| Observations | 54,404 | 54,404 | 54,404 | 54,404 | 54,404 |

Notes: All estimations are based on the GSCM method. Robust standard errors clustered at project level in parentheses. *** p<0.01, ** p<0.05, * p<0.1.

Table J.2: The Impact of Copilot on Complex Projects

| | (1)<br>Project-level Code Contributions | (2)<br>Coordination Time | (3)<br>Timely Project Outputs (1 Day) | (4)<br>Timely Project Outputs (3 Days) | (5)<br>Timely Project Outputs (10 Days) |
|---|---|---|---|---|---|
| | Log (Merged_PR) | Log (Merge_time) | Log (MergedPR_1D) | Log (MergedPR_3D) | Log (MergedPR_10D) |
| ATT of Copilot | 0.086*** | 0.068** | 0.065*** | 0.071*** | 0.079*** |
| | (0.012) | (0.029) | (0.014) | (0.013) | (0.014) |
| Project FE | Yes | Yes | Yes | Yes | Yes |
| Month FE | Yes | Yes | Yes | Yes | Yes |
| # of projects | 4,046 | 4,046 | 4,046 | 4,046 | 4,046 |
| Observations | 78,706 | 78,706 | 78,706 | 78,706 | 78,706 |

Notes: All estimations are based on the GSCM method. Robust standard errors clustered at project level in parentheses. *** p<0.01, ** p<0.05, * p<0.1.

Table J.3: The Impact of Copilot on Project Complexity (PSM and DID)

| | (1)<br>Project-level Code Contributions | (2)<br>Coordination Time |
|---|---|---|
| | Log (Merged_PR) | Log (Merge_time) |
| Copilot | 0.142*** | 0.150*** |
| | (0.034) | (0.057) |
| Copilot*Complexity | 0.059 | -0.003 |
| | (0.039) | (0.063) |
| Project FE | Yes | Yes |
| Month FE | Yes | Yes |
| # of projects | 2,122 | 2,122 |
| Observations | 42,251 | 42,251 |

Notes: Robust standard errors clustered at project level in parentheses. *** p<0.01, ** p<0.05, * p<0.1.

## Appendix K: Additional Analyses on the Effects of Copilot on Core and Peripheral Developers

We construct a set of metrics to capture absolute changes in project-level outcomes for both core and peripheral developers. These metrics span four key dimensions: project-level code contributions, coordination time, project-level timely merge of code contributions, and the underlying mechanisms including developer coding participation, individual productivity, code discussion volume, code discussion participation, and code discussion intensity per developer.

Specifically, for each project-month, we compute the following variables for core and peripheral developers: (1) the number of merged PRs submitted by each group; (2) the average time taken to merge a single PR submitted by each group; (3) the number of PRs submitted by each group that were merged within one, three, and ten days; (4) the number of distinct core and peripheral developers submitting PRs which were eventually merged; (5) the average number of merged PRs per core and peripheral developer; (6) the average number of comments per merged PR submitted by each group; (7) the average number of unique developers participating in code discussions per merged PR submitted by each group; and (8) the average number of comments per developer per merged PR submitted by each group.

We report the results for core developers' project-level code contributions, coordination time, and project-level timely merge of code contributions in Table K.1 and for peripheral developers in Table K.2. Overall, the findings indicate that both core and peripheral developers benefited from the adoption of Copilot, though the magnitude and nature of these benefits varied. For core developers, Copilot led to a 6.1% increase in project-level code contributions, a 6.7% increase in coordination time, and increases in timely integrated code contributions within one, three, and ten days by 4.4%, 4.8%, and 5.8%, respectively. In contrast, peripheral developers experienced a 5.4% increase in project-level code contributions and a 5.3% increase in coordination time. Their timely integrated code contributions increased by 2.3%, 2.4%, and 3.6% within one, three, and ten days, respectively.

The results for the underlying mechanisms, such as developer coding participation, individual code contributions, and code discussion-related metrics, are presented in Table K.3 for core developers and Table K.4 for peripheral developers. As shown in Table K.3, core developers experienced a 5.5% increase in developer coding participation, a 5.6% increase in individual code contributions, and increases of 4.1%, 1.6%, and 3.6% in code discussion volume, code discussion participation, and code discussion intensity per developer, respectively.

For peripheral developers, as indicated in Table K.4, they experienced a 2.8% increase in developer coding participation, a 2.2% increase in individual code contributions, and increases of 12.6%, 2.3%, and 11.4% in code discussion volume, code discussion participation, and code discussion intensity per developer, respectively.

Table K.1: The Effect of Copilot on Core Developers (Absolute Changes)

| | (1) Project-level Code Contributions | (2) Coordination Time | (3) Timely Merged PRs (1 Day) | (4) Timely Merged PRs (3 Days) | (5) Timely Merged PRs (10 Days) |
|---|---|---|---|---|---|
| | Log (Core_ merged_PR) | Log (Core_ merge_time) | Log (Core_ mergedPR_1D) | Log (Core_ mergedPR_3D) | Log (Core_ mergedPR_10D) |
| ATT of Copilot | 0.061*** | 0.067*** | 0.044*** | 0.048*** | 0.058*** |
| | (0.008) | (0.023) | (0.011) | (0.010) | (0.010) |
| Project FE | Yes | Yes | Yes | Yes | Yes |
| Month FE | Yes | Yes | Yes | Yes | Yes |
| # of projects | 6,423 | 6,423 | 6,423 | 6,423 | 6,423 |
| Observations | 118,853 | 118,853 | 118,853 | 118,853 | 118,853 |

Notes: All estimations are based on the GSCM method. Robust standard errors clustered at project level in parentheses. *** p<0.01, ** p<0.05, * p<0.1.

Table K.2: The Effect of Copilot on Peripheral Developers (Absolute Changes)

| | (1) Project-level Code Contributions | (2) Coordination Time | (3) Timely Merged PRs (1 Day) | (4) Timely Merged PRs (3 Days) | (5) Timely Merged PRs (10 Days) |
|---|---|---|---|---|---|
| | Log (Peri_ merged_PR) | Log (Peri_ merge_time) | Log (Peri_ mergedPR_1D) | Log (Peri_ mergedPR_3D) | Log (Peri_ mergedPR_10D) |
| ATT of Copilot | 0.054*** | 0.053* | 0.023** | 0.024** | 0.036*** |
| | (0.008) | (0.028) | (0.010) | (0.012) | (0.010) |
| Project FE | Yes | Yes | Yes | Yes | Yes |
| Month FE | Yes | Yes | Yes | Yes | Yes |
| # of projects | 4,154 | 4,154 | 4,154 | 4,154 | 4,154 |
| Observations | 77,888 | 77,888 | 77,888 | 77,888 | 77,888 |

Notes: All estimations are based on the GSCM method. Robust standard errors clustered at project level in parentheses. *** p<0.01, ** p<0.05, * p<0.1.

Table K.3: The Effect of Copilot on Core Developers (Absolute Changes, Mechanism Analyses)

| | (1)<br>Coding Participation | (2)<br>Individual Code Contributions | (3)<br>Discussion Volume | (4)<br>Discussion Participation | (5)<br>Individual Discussion Intensity |
|---|---|---|---|---|---|
| | Log (Num_core_with_mergedPR) | Log (MergedPR_per_core) | Log (Comment_per_mergedPR) | Log (Num_dev_with_comments_per_mergedPR) | Log (Comments_per_dev_per_mergedPR) |
| ATT of Copilot | 0.055*** | 0.056*** | 0.041** | 0.016*** | 0.036** |
| | (0.007) | (0.011) | (0.019) | (0.005) | (0.017) |
| Project FE | Yes | Yes | Yes | Yes | Yes |
| Month FE | Yes | Yes | Yes | Yes | Yes |
| # of projects | 6,423 | 6,423 | 6,423 | 6,423 | 6,423 |
| Observations | 118,853 | 118,853 | 118,853 | 118,853 | 118,853 |

Notes: All estimations are based on the GSCM method. Robust standard errors clustered at project level in parentheses. *** p<0.01, ** p<0.05, * p<0.1.

Table K.4: The Effect of Copilot on Peripheral Developers (Absolute Changes, Mechanism Analyses)

| | (1)<br>Coding Participation | (2)<br>Individual Code Contributions | (3)<br>Discussion Volume | (4)<br>Discussion Participation | (5)<br>Individual Discussion Intensity |
|---|---|---|---|---|---|
| | Log (Num_peri_with_mergedPR) | Log (MergedPR_per_peri) | Log (Comment_per_mergedPR) | Log (Num_dev_with_comments_per_mergedPR) | Log (Comments_per_dev_per_mergedPR) |
| ATT of Copilot | 0.028*** | 0.022*** | 0.126*** | 0.023*** | 0.114*** |
| | (0.005) | (0.006) | (0.033) | (0.007) | (0.026) |
| Project FE | Yes | Yes | Yes | Yes | Yes |
| Month FE | Yes | Yes | Yes | Yes | Yes |
| # of projects | 4,154 | 4,154 | 4,154 | 4,154 | 4,154 |
| Observations | 77,888 | 77,888 | 77,888 | 77,888 | 77,888 |

Notes: All estimations are based on the GSCM method. Robust standard errors clustered at project level in parentheses. *** p<0.01, ** p<0.05, * p<0.1.

In addition, we examine whether the increase in participation is more pronounced among new or existing peripheral developers. We classify peripheral developers as new or existing based on whether they have previously contributed to the focal project. Following the ratio measure used in the main analyses, we calculate the proportion of new peripheral developers relative to the total number of peripheral developers and determine the effect of Copilot on this variable. A positive estimate indicates a greater increase in the participation of new peripheral developers, compared to existing peripheral developers.

As shown in column (1) of Table K.5 below, there was a significant increase in the share of new peripheral developers following the use of Copilot. Moreover, column (2) of Table K.5 shows a significant increase in new peripheral developer participation as well, suggesting that the observed increase in the share of new peripheral developers was not driven by a decrease in existing developer participation.

Overall, this pattern seems consistent with our cost–benefit argument. For peripheral developers, the expected benefits of contributing to the focal project are relatively stable, as their involvement is typically limited in scope and influence. The introduction of AI pair programmers primarily affects the cost side of the participation decision for them. For new peripheral developers, AI substantially lowers expected costs, thereby increasing the net value of contribution and encouraging participation. In contrast, existing peripheral developers appear to experience smaller cost reductions. With similar expected benefits, the more modest cost savings translate into a relatively limited increase in net contribution value, which weakens their incentive to further increase participation.

Table K.5: The Effect of Copilot on New and Existing Peripheral Developers Participation

| | (1)<br>New_dev_with_mergedPR_pct | (2)<br>Log (nums_new_dev_with_mergedPR) |
|---|---|---|
| ATT of Copilot | 0.039*** | 0.094*** |
| | (0.003) | (0.007) |
| Project FE | Yes | Yes |
| Month FE | Yes | Yes |
| # of projects | 3,584 | 3,584 |
| Observations | 84,807 | 84,807 |

Notes: All estimations are based on the GSCM method. Robust standard errors clustered at project level in parentheses. *** p<0.01, ** p<0.05, * p<0.1.

## Appendix L: Project Familiarity

We argue that one potential reason that explains the difference in code contributions and coordination time between core and peripheral developers lies in their different levels of project familiarity. In the OSS community, core developers are more embedded in focal projects and maintain sustained involvement (Setia et al. 2012), while peripheral developers contribute more sporadically (Howison and Crowston 2014, Krishnamurthy et al. 2016), which leads to their limited level of project familiarity.

To explore this potential explanation, we first compare the level of project familiarity for core versus peripheral developers. Based on the sample of projects with continuous activity from 2018 to 2021, we measure a developer's project familiarity based on their activity during this period related to a focal project. More specifically, a developer's project familiarity with a focal project is measured by the developer's working tenure with the project, i.e., the number of months a developer contributed to the focal project before June 2021 (i.e., the introduction of GitHub Copilot). As reported in Table L.1, the t-test results show that on average, core developers have significantly longer working tenure than peripheral developers, consistent with our argument that core developers have greater project familiarity.

Table L.1: Comparison between Core and Peripheral Developers on their Project Familiarity, Measured by Project Working Tenure

| Variable | Core Developers | Peripheral Developers | Mean diff | p-value |
|---|---|---|---|---|
| Avg_tenure | 6.98 | 1.64 | 5.34 | 0.00 |

Notes: The table is based on the same sample as the one used in our main analyses in the paper.

We next explore whether the differential effects of Copilot on code contributions and coordination time among core versus peripheral developers can be partly explained by the different levels of project familiarity. To do so, our main idea is to examine whether Copilot had differential effects on code contributions and coordination time for developers with different levels of familiarity. Developers are classified into high versus low project familiarity based on whether their working tenure exceeds the median value of working tenure for that project before the adoption of Copilot. Using the same empirical approach as Section 6.5, we compute the ratio of code contributions made by high familiarity developers, denoted as the *Ratio of merged PR*. We also compute the coordination time required to merge their contributions relative to the project average, denoted as the *Ratio of merge time*. The results are shown in Tables L.2. Consistent with our arguments, Copilot adoption led to a relatively greater increase in the share of code contributions made by developers with high project familiarity, alongside a relatively greater reduction in the coordination time required to merge their contributions compared to the project level average.

Table L.2: Heterogeneous Effect of Copilot on Developers with High vs. Low Project Familiarity

| | (1)<br>Project-level Code Contributions | (2)<br>Overall Coordination Time |
|---|---|---|
| | Ratio of merged PR | Ratio of merge time |
| ATT of Copilot | 0.024*** | -0.052*** |
| | (0.004) | (0.009) |
| Project FE | Yes | Yes |
| Month FE | Yes | Yes |
| # of projects | 6,171 | 6,171 |
| Observations | 87,048 | 87,048 |

Notes: All estimations are based on the GSCM method. Robust standard errors clustered at project level in parentheses. *** p<0.01, ** p<0.05, * p<0.1.

Furthermore, we analyze the content of discussions surrounding both core and peripheral developers' code contributions. We find that discussions related to peripheral developers' contributions frequently focus on clarifying design intent, aligning code with project specific context and historical development, and resolving uncertainties related to integration risks. Such comments are consistent with more limited project familiarity and contextual knowledge, which in turn leads to more discussions during code integration. We provide representative qualitative examples below to illustrate these patterns:

*"You can find the table of valid contexts here."*

*"That's probably better (more semantically accurate) in the context of this project."*

*""configuring" got me a bit confused, since it sounds like a process, I couldn't figure out which point in time exactly it is."*

*"I think this changes the value of the global variable XXX if it succeeds"*

*"I added more context -- let's discuss more here. This is a very risky area of the code for compatibility."*

*"I'm not sure I do understand the intended flow here. My understanding is XXX. If this is the case, you should rather do something like XXX."*

*"We would like to avoid using XXX in all other places. So, how about we change this as below?"*

To assess whether these qualitative examples are representative, we analyze the content of PR discussions. Specifically, we employ a dictionary-based text analysis approach, which is particularly feasible given the scale of our data, comprising 5,404,735 discussion records. We apply keyword matching to PR comments associated with code contributions by core and peripheral developers. Specifically, we capture context clarification discussion, including comments aimed at clarifying design intent, aligning code with project specific context and historical development, and resolving integration related uncertainties. Prior work documents that such clarification-oriented feedback is more prevalent when contributors have more limited project familiarity (Rullani and Haefliger 2013, Tsay et al. 2014, Joblin et al. 2017).

We calculate the proportion of comments received by core and peripheral developers that relate to context clarification and report these results in Table L.3. The results show that peripheral developers receive a higher proportion of comments related to context clarification, which suggests that they are less familiar with the project context.

Overall, this set of evidence provides some additional support that project familiarity might partially explain the difference in the impact of Copilot on coordination time between core developers and peripheral developers.

Table L.3: Comparison between Core and Peripheral Developers on their Received Comment

| Variable | Core Developers | Peripheral Developers | Mean diff | p-value |
|---|---|---|---|---|
| Context_comment_pct | 0.087 | 0.204 | -0.117 | 0.00 |

**Appendix M: Alternative Explanations for the Differential Impacts between Core and Peripheral Developers**

To address the concern that the heterogeneous effects of Copilot on core versus peripheral developers may be driven by differences such as programming language expertise or Copilot usage, we conduct the following additional analyses.

First, it is worth noting that the theoretical distinction between core and peripheral developers does not necessarily reflect differences in programming language expertise. As suggested by the literature, peripheral developers include core developers from other projects and end users (Setia et al. 2012), which indicates that peripheral developers may also possess substantial expertise in the relevant programming languages. Core developers may be new to a domain and rely on peripheral contributors with specialized knowledge to support project development. Alternatively, core developers may be domain experts, while peripheral developers participate in the project to learn and gain experience.

Given the plausibility of both scenarios, we compare the fraction of expert developers in core developers against the fraction of expert developers in peripheral developers. Specifically, we classify developers as programming language experts if they used that language in at least 75% of their prior projects. We then compare each developer's expertise to the primary language used in the focal project to determine whether they met the expert criteria.

As reported in the first row in Table M.1, the t-test result shows no significant difference between core and peripheral developers in terms of the percentage of expert developers. This suggests that differences in language expertise may not be the primary reason for the observed heterogeneity between core and peripheral contributors.

Second, we consider the possibility that core developers benefit more simply because they are more likely to use Copilot or use it more intensively. To explore this possibility, we leverage the proprietary data provided by GitHub organization that indicates for each project, the percentage of Copilot users among core developers and the percentage of Copilot users among peripheral developers. Due to privacy constraints, we do not have access to individual-level Copilot usage data for specific core or peripheral developers.

As reported in the second row in Table M.1, interestingly, we find that peripheral developers had a significantly higher Copilot adoption rate than core developers. Despite this, our analysis in Section 6.5 of the main paper shows that peripheral developers experienced relatively smaller project-level code contributions and greater coordination time from Copilot usage. This pattern further supports our argument that the observed differences in outcomes are not driven by differential Copilot adoption rates but are more plausibly attributed to differences in project familiarity.

Table M.1: Comparison between Core and Peripheral Developers

| Variable | Core Developers | Peripheral Developers | Mean diff | p-value |
|---|---|---|---|---|
| % of expert developers | 0.68 | 0.67 | 0.01 | 0.15 |
| % of Copilot users | 0.20 | 0.30 | -0.10 | 0.00 |

Notes: The table is based on the same sample as the one used in our main analyses in the paper.

To further address the concern that increased activity by core developers may crowd out contributions from peripheral developers, leading to the cannibalization effect, we include additional controls that capture prior coordination time within the project. Specifically, we control for the lagged log average merge time of core developers' code and the lagged log average merge time of peripheral developers' code. As reported in Table M.2, the results remain consistent after including the controls. That is, we continue to observe an increase in the number of peripheral contributions and a decline in the proportion of peripheral contributions relative to total contributions following Copilot adoption.

Table M.2: The Effect of Copilot on Peripheral Developers' Code Contributions

| | (1)<br>Log (Peri_merged_PR) | (2)<br>Peri_merged_PR_share |
|---|---|---|
| ATT of Copilot | 0.081*** | -0.022*** |
| | (0.018) | (0.007) |
| Log (avg_core_merge_time_lag) | -0.009** | -0.006*** |
| | (0.004) | (0.002) |
| Log (avg_peri_merge_time_lag) | -0.0002 | 0.004** |
| | (0.004) | (0.002) |
| Project FE | Yes | Yes |
| Month FE | Yes | Yes |
| # of projects | 4,154 | 4,154 |
| Observations | 77,888 | 77,888 |

Notes: All estimations are based on the GSCM method. Robust standard errors clustered at project level in parentheses. *** p<0.01, ** p<0.05, * p<0.1.

To address the review priority explanation that peripheral developers may wait longer for their code to be reviewed (Yu et al. 2015), we measure coordination time as the duration from the first review to final acceptance, thereby excluding any waiting time prior to the initial review. As shown in Table M.3, both core and peripheral developers experienced an increase in this review duration, consistent with the rise in code discussion reported in Tables K.3 and K.4 of Online Appendix K. Importantly, peripheral developers exhibited a larger increase in this duration. Overall, the results mitigate concerns about review priority, as this alternative measure isolates coordination during the review process rather than delays in obtaining an initial review.

Table M.3: The Effect of Copilot on Review Duration for Core and Peripheral Developers

| | (1) Core Developers' Review Duration | (2) Peripheral Developers' Review Duration | (3) Peripheral Developers' Review Duration Relative to Core Developers |
|---|---|---|---|
| | Log (Core_review_to_merge) | Log (Peri_review_to_merge) | Peri_review_to_merge_ratio |
| ATT of Copilot | 0.074*** | 0.115*** | 0.049*** |
| | (0.015) | (0.014) | (0.016) |
| Project FE | Yes | Yes | Yes |
| Month FE | Yes | Yes | Yes |
| # of projects | 6,423 | 4,154 | 4,154 |
| Observations | 118,853 | 77,888 | 77,888 |

Notes: All estimations are based on the GSCM method. Robust standard errors clustered at project level in parentheses. *** p<0.01, ** p<0.05, * p<0.1.